\documentclass[twocolumn,english,prb,aps,showpacs,floatfix,groupedaddress,superscriptaddress,tightenlines]{revtex4}
\usepackage[]{fontenc}
\usepackage[latin9]{inputenc}
\usepackage{amsmath}
\usepackage{graphicx}
\usepackage{amssymb}

\makeatletter

\newcommand{\lyxdot}{.}

\@ifundefined{textcolor}{}
{%
 \definecolor{BLACK}{gray}{0}
 \definecolor{WHITE}{gray}{1}
 \definecolor{RED}{rgb}{1,0,0}
 \definecolor{GREEN}{rgb}{0,1,0}
 \definecolor{BLUE}{rgb}{0,0,1}
 \definecolor{CYAN}{cmyk}{1,0,0,0}
 \definecolor{MAGENTA}{cmyk}{0,1,0,0}
 \definecolor{YELLOW}{cmyk}{0,0,1,0}
 }

\newcommand{\hc}{\mathrm{h.c.}}\newcommand{\tr}{\mathrm{tr}}

\newcommand{\1}{\leavevmode{\rm 1\ifmmode\mkern  -4.8mu\else\kern -.3em\fi I}}
\makeatother

\makeatother

\usepackage{babel}

\makeatother

\usepackage{babel}

\makeatother

\usepackage{babel}

\begin{document}

\title{Excitation transfer through open quantum networks: a few basic mechanisms}

\author{Lorenzo Campos Venuti}

\affiliation{Institute for Scientific Interchange (ISI), Viale S. Severo 65, I-10133
Torino, Italy }

\author{Paolo Zanardi}

\affiliation{Department of Physics and Astronomy and Center for Quantum Information
Science \& Technology, University of Southern California, Los Angeles,
California 90089-0484, USA}

\affiliation{Institute for Scientific Interchange (ISI), Viale S. Severo 65, I-10133
Torino, Italy }
\begin{abstract}
A variety of open quantum networks are currently under intense examination
to model energy transport in photosynthetic systems. Here we study
the coherent transfer of a quantum excitation over a network incoherently
coupled with a structured and small environment that effectively models
the photosynthetic reaction center. Our goal is to distill a few basic,
possibly universal, mechanisms or {}``effects\char`\"{} that are
featured in simple energy-transfer models. In particular, we identify
three different phenomena: the congestion effect, the asymptotic unitarity
and the staircase effects. We begin with few-site models, in which
these effects can be fully understood, and then proceed to study more
complex networks similar to those employed to model energy transfer
in light-harvesting complexes. Our numerical studies on such networks
seem to suggest that some of the effects observed in simple networks
may be of relevance for biological systems, or artificial analogues
of them as well. 
\end{abstract}
\maketitle

\section{Introduction}

The transport of electronic excitations over biological networks of
chromophores is the relevant mechanism for the light-harvesting step
of photosynthesis \cite{BlankenshipBook,vanGrondelle,VanGrondelle2,Renger2006,Renger2009,RitzSchulten2002}.
Recently, long-lived quantum coherent oscillations have been observed
in ultrafast experiments carried out on several biological systems,
even at room temperature \cite{Fleming2005,Engel2007,Scholes2010,Engel2010,Fleming2010,AndrewMoran2010JCP}.
One of the key features of these exciton-transfer networks is their
open nature, namely, that their coupling with the protein vibrational
environment is, arguably, the dominant effector of transport in these
systems. The interplay of unitary dynamics and the system-bath interaction
has been predicted to be beneficial to the network functionality at
biological conditions \cite{GaabBardeen2004,Mohseni2008,Rebentrost2009,Rebentrost2009b,PlenioHuelga2008,PlenioHuelga2009,CaoSilbey2009,CaoSilbey2010,ENAQTDimersMuelken2010arxiv}.
The different mechanisms that lead to this \textsl{environment-assisted
quantum transport} \cite{Rebentrost2009} are still under vigorous
exploration \cite{CaoSilbey2010}. Realistic numerical modeling of
these open quantum networks is, to some extent, possible and currently
actively pursued in the physical chemistry community \cite{Cao1997,MarcusRenger,OlayaCastro2008,Ishizaki2009JCP,CaoSilbey2010,MohseniLloyd2010,KaisAspuru2010,Nazir,Thorwart,OlayaCastro2010,OlayaCastro2010b}.
Nevertheless, the physical chemistry and quantum information community
has learned much from simple Markovian models \cite{OlayaCastro2008,Mohseni2008,OlayaCastro2010}.


In this paper, motivated by the above, we will investigate a few simple
yet illuminating models of open quantum networks in order to identify
a handful of basic mechanisms or effects that are featured in fully
analyzable \textsl{toy models} and that may persist for larger, more
complex quantum transport networks. In particular, we will focus on
coherently-coupled qubits subject to dissipation/dephasing and irreversibly
connected to an auxiliary quantum system. The role of this latter
is to model the reaction center of light-harvesting complexes, where
the electronic excitation is separated into an electron and a hole
and the charge-transfer stage of photosynthesis begins. Of interest
to us is the reaction center of the LH1-RC complexes present in purple
bacteria \cite{OlayaCastro2008,RitzSchulten2002}. We will adopt a
Markovian master equation of the Lindblad form to describe the overall
system dynamics. Different energies, or equivalently time-scales,
will enter the definition of the Liouville superoperator $\mathcal{L}$.
The interplay of these time-scales controls the non-trivial phenomenology
that we explore in this manuscript. Finally, singling out a few intriguing,
possibly universal features of such a phenomenological landscape is
the goal of the simple calculations presented in this paper.

In the next three sections (\ref{sec:congestion}, III and IV) we
will consider different toy models consisting of few sites or chromophores
(modeled as quantum two-level systems, or qubits), manifesting particular
features which can be fully understood by analytical calculations.
See Fig.~\ref{fig:toy_networks} for a cartoon picture of the various
networks considered. In section \ref{sec:Applications-to-LH1-RC},
we will consider more realistic networks borrowed from models of light-harvesting
complexes. Via numerical simulations we will show that these effects
may persist in more realistic systems.

\section{The congestion effect\label{sec:congestion} }

In exciton and electron transfer events, there can be delays in energy
transport due to the timescales of the biological process. A particular
element might be shut down while transport takes place, effectively
making an exciton or electron wait until the transport is possible
\cite{BatteryPaper2010}. In the following section, we will describe
this phenomenon in model systems and characterize it as the \textsl{congestion
effect}.

In the standard modeling of incoherent (and irreversible) transfer
of excitations from one site to another, the Förster electromagnetic
coupling mechanism permits the transfer of populations at a given
rate $\gamma$. If the dynamics is described using a Lindblad form
$\dot{\rho}=\mathcal{L}_{L}\left(\rho\right),$ where $\mathcal{L}_{L}\left(X\right)=LXL^{\dagger}-\left\{ L^{\dagger}L,X\right\} /2$,
this can be accounted for by a jump operator of the form $L=\sqrt{\gamma}\sigma^{-}\otimes\sigma^{+},$
where $\sigma^{\pm}$ are Pauli ladder operators. For our purposes,
such a Lindblad description is phenomenological. Site 2 could model,
for example the reaction center of LH-II described above. 
In this section, we explore possible \textsl{congestion} effects that
arise from the dependence of the transfer rate on the number of excitations
involved, in the same way traffic flow might be inversely proportional
to the number of vehicles present on roads.

\begin{figure}
\noindent \begin{centering}
\includegraphics[width=6.5cm]{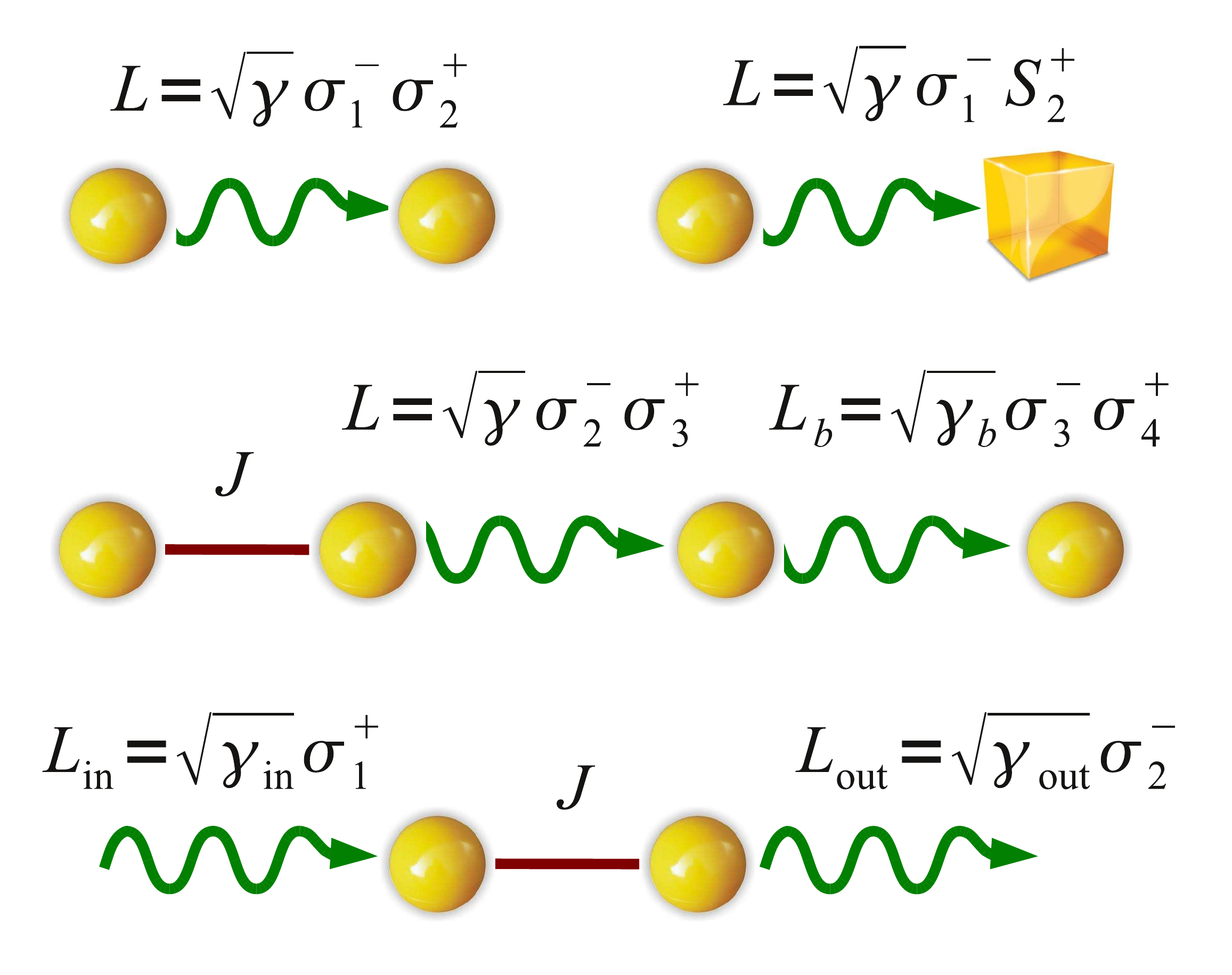} 
\par\end{centering}

\caption{Summary of the toy-networks analyzed analytically in sections \ref{sec:congestion}
and \ref{sec:staircase}.\label{fig:toy_networks}}

\end{figure}

\paragraph*{Incoherent transfer I: $\bullet\stackrel{\gamma}{\rightsquigarrow}\bullet$.}

Before turning to analyze the possible implementations and consequences
of such an effect, let us summarize the Lindblad operators for incoherent
Förster transfer among two sites, $L=\sqrt{\gamma}\sigma_{1}^{-}\sigma_{2}^{+}$.
This process can be pictorially visualized by the following diagram,
$\bullet\stackrel{\gamma}{\rightsquigarrow}\bullet$ (see also Fig.~\ref{fig:toy_networks}).
The quantum master equation is given simply by $\dot{\rho}=\mathcal{L}_{L}\left(\rho\right)$.
We denote by $\mathfrak{n}$ the population operator satisfying $\mathfrak{n}|\eta\rangle=\eta|\eta\rangle$
with $\eta=0,1$, and by $n$ its possibly time-dependent expectation
value for excitations, i.e.~$n=\langle\mathfrak{n}\rangle:=\tr\left(\mathfrak{n}\rho\right)$.
Since the effect of the Lindbladian is to transfer a particle from
site $1$ to site $2$, the total number operator is a conserved quantity.
We therefore obtain a differential equation for the population in
the following way: first note that $\dot{n}_{i}=\tr\left(\mathfrak{n}_{i}\dot{\rho}\right)=\tr\left[\mathfrak{n}_{i}\mathcal{L}_{L}\left(\rho\right)\right].$
Given that $n_{1}+n_{2}=n_{\mathrm{tot}}$ is constant in time, it
suffices to analyze the population of site 1, $\dot{n}_{1}=-\gamma n_{1}+\gamma\langle\mathfrak{n}_{1}\mathfrak{n}_{2}\rangle$.
Now note that in the single-particle sector, $\mathfrak{n}_{\mathrm{tot}}=1$,
$\langle\mathfrak{n}_{1}\mathfrak{n}_{2}\rangle=0$ (to see this use
$\mathfrak{n}_{\mathrm{tot}}^{2}=\mathfrak{n}_{\mathrm{tot}}+2\mathfrak{n}_{1}\mathfrak{n}_{2}$),
leading to a transport equation $\dot{n}_{1}=-\gamma n_{1}$ that
can be readily solved for the population at sites 1, $n_{1}\left(t\right)=e^{-\gamma t}n_{1}\left(0\right)$
and 2, $n_{2}\left(t\right)=n_{2}\left(0\right)+\left(1-e^{-\gamma t}\right)n_{1}\left(0\right)$.
The jump operator achieves precisely what we expected: the population
in site one decreases exponentially at a rate $\gamma$ and the population
of site 2 increases accordingly. The same result could have been obtained
by solving the (16 dimensional) differential equation for the full
density matrix. Starting at time zero with $\rho\left(0\right)=\{\rho_{i,j}\}$
the time-evolved density matrix $\rho\left(t\right)$ is\[
\left(\begin{array}{cccc}
\rho_{1,1} & e^{-\gamma t/2}\rho_{1,2} & \rho_{1,3} & \rho_{1,4}\\
e^{-\gamma t/2}\rho_{2,1} & e^{-\gamma t}\rho_{2,2} & e^{-\gamma t/2}\rho_{2,3} & e^{-\gamma t/2}\rho_{2,4}\\
\rho_{3,1} & e^{-\gamma t/2}\rho_{3,2} & \left(1-e^{-\gamma t}\right)\rho_{2,2}+\rho_{3,3} & \rho_{3,4}\\
\rho_{4,1} & e^{-\gamma t/2}\rho_{4,2} & \rho_{4,3} & \rho_{4,4}\end{array}\right).\]
 It is interesting to note that for some entangled initial states
the asymptotic density matrix $\rho\left(t\rightarrow\infty\right)$
is still entangled. The process $\mathcal{L}_{L}$ cannot, however,
create entanglement.

\paragraph*{Incoherent transfer II: $\bullet\rightsquigarrow\square$.}

To model the congestion in the reaction center, let us now substitute
the second qubit with a larger $2s+1$ dimensional space. 

For this case, we can model a particle conserving transfer process
with a jump operator given by $L=\sqrt{\gamma}\sigma_{1}^{-}S_{2}^{+}$
where $S_{2}^{+}$ is a raising operator of the irreducible spin $s$
representation of $SU\left(2\right)$. The population at site 2 is
$\mathfrak{N}_{2}=S_{2}^{z}+s\1$. Once again, since the total particle
number $\mathfrak{n}_{\mathrm{tot}}=\mathfrak{n}_{1}+\mathfrak{N}_{2}$
is conserved in a given particle sector, one has $\mathfrak{n}_{\mathrm{tot}}\left(t\right)=n_{\mathrm{tot}}$.
We then obtain the following differential equation for population
at site 1: $\dot{n}_{1}=-\gamma\langle\mathfrak{n}_{1}S_{2}^{-}S_{2}^{+}\rangle$.
By noting that $S_{2}^{-}S_{2}^{+}=(\mathfrak{N}_{2}+1)(2s-\mathfrak{N}_{2})$,
and employing $\mathfrak{N}_{2}=n_{\mathrm{tot}}-\mathfrak{n}_{1}$,
$\mathfrak{N}_{2}^{2}=n_{\mathrm{tot}}^{2}-2n_{\mathrm{tot}}-1+2\mathfrak{n}_{1}$,
and $\mathfrak{n}_{1}^{2}=\mathfrak{n}_{1}$, we obtain an explicit
differential equation for $n_{1}$: \begin{eqnarray*}
\dot{n}_{1} & = & -\gamma n_{\mathrm{tot}}\left[(2s+1)-n_{\mathrm{tot}}\right]\, n_{1}\\
n_{1}+N_{2} & = & n_{\mathrm{tot}}.\end{eqnarray*}

Excitation transfer now occurs at an effective rate which depends
on the total population: $\gamma_{\mathrm{eff}}=\gamma n_{\mathrm{tot}}\left[(2s+1)-n_{\mathrm{tot}}\right]$.
Note that $0\le n_{\mathrm{tot}}\le2s+1$ and, correctly, $\gamma_{\mathrm{eff}}\left(n_{\mathrm{tot}}=0\right)=\gamma_{\mathrm{eff}}\left(n_{\mathrm{tot}}=2s+1\right)=0$,
i.e.~no transfer takes place when the network is either completely
empty or completely full. The maximum transfer rate is attained when
the condition $n_{\mathrm{tot}}=\left(2s+1\right)/2$ is satisfied.
The lesson we get from this slightly modified example, is that transferring
excitations to an object with more than just two levels, is likely
to result in a population dependent transfer rate.

\paragraph*{Interplay between coherent hopping and transfer: $\bullet\stackrel{J}{\leftrightarrow}\bullet\stackrel{\gamma}{\rightsquigarrow}\bullet\stackrel{\gamma_{b}}{\rightsquigarrow}\bullet$.}

\label{sec:hoppingtransfer} %
\begin{figure}
\noindent \begin{centering}
\includegraphics[width=6.5cm]{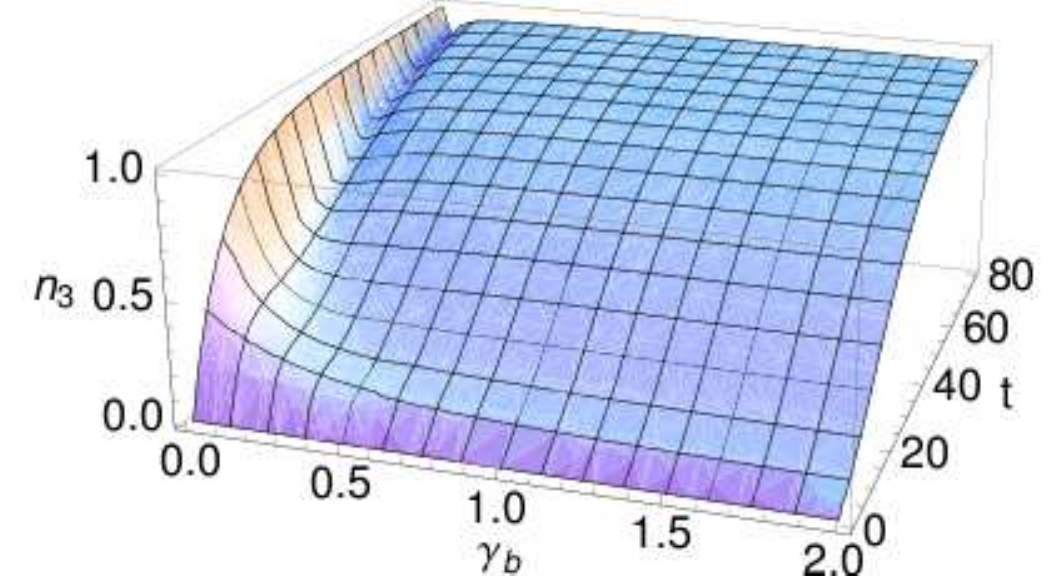} 
\par\end{centering}

\noindent \begin{centering}
\includegraphics[width=6.5cm]{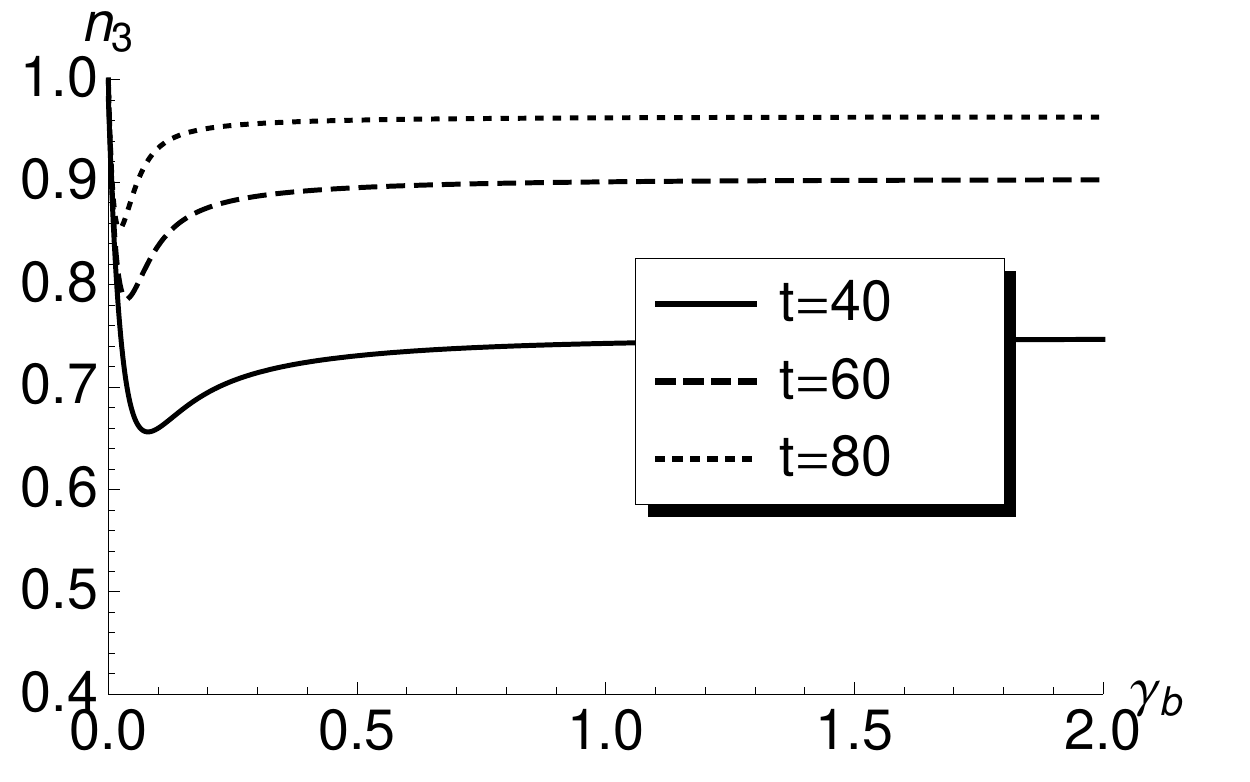} 
\par\end{centering}

\caption{\textsl{Top panel:} The population $n_{3}$ for the case described
in section \ref{sec:hoppingtransfer} , as a function of time and
$\gamma_{b}$. In this model, the initial state has two excitations
at sites 1 and 2: $|1,1,0,0\rangle$. The parameters for the model
are $J=1,\,\gamma=0.1$. \textsl{Bottom: panel}: Slices of the same
plot at different times are shown. The non-monotonic behavior of the
population as a function of the rate $\gamma_{b}$ is evident at small
values of it. \label{fig:n3_2exc}}

\end{figure}

We will further illustrate the concept above by considering a variation
on the theme. We consider a coherent-hopping Hamiltonian on four sites
of the form $H=\left(J/2\right)\left(\sigma_{1}^{-}\sigma_{2}^{+}+\hc\right)$
that acts on the first two sites. The excitations are transferred
irreversibly from site 2 to site 3 via a quantum jump operator $L=\sqrt{\gamma}\sigma_{2}^{-}\sigma_{3}^{+}$
and subsequently from site 3 to site 4 with $L_{b}=\sqrt{\gamma_{b}}\sigma_{3}^{-}\sigma_{4}^{+}$.
$J$ is the coherent coupling strength. In the following, we explore
the interplay between the two incoherent transfer rates $\gamma$
and $\gamma_{b}$. Let us focus on the population at site 3, $n_{3}\left(t\right)$.
The effect of $\gamma_{b}$ is that of removing excitation population
from site 3. However when $\gamma_{b}$ becomes large, excitations
are rapidly transferred to site 4 inhibiting the effect of $L_{b}$
(\emph{$\mathcal{L}_{b}\left(\rho\right)\rightarrow0$}). This results
in a non-trivial non-monotonic effect as a function of $\gamma_{b}$.
This feature can be visible only if we have at least two particles
in the network. Let us then consider the following initial (pure)
state with excitations localized at sites 1 and 2: $|1,1,0,0\rangle$.
As shown in Figure 3, in this case, the time-evolution of the populations
takes the following form:

\begin{eqnarray*}
n_{1}\left(t\right) & = & C_{1}e^{-\gamma t}+C_{2}e^{-\gamma t/2}+C_{3}\left(t\right)e^{-t\gamma_{b}}\\
 &  & +C_{4}e^{-t\left(\gamma+\omega\right)/2}+C_{5}e^{-t\left(\gamma-\omega\right)/2}\\
n_{2}\left(t\right) & = & C'_{1}e^{-\gamma t}+C'_{2}e^{-\gamma t/2}+C'_{3}\left(t\right)e^{-t\gamma_{b}}\\
 &  & +C'_{4}e^{-t\left(\gamma+\omega\right)/2}+C'_{5}e^{-t\left(\gamma-\omega\right)/2}\\
n_{3}\left(t\right) & = & 1+B_{1}\left(t\right)e^{-t\gamma_{b}}+B_{2}e^{-t\gamma}+B_{3}e^{-t\gamma/2}\\
 &  & +B_{4}e^{-t\left(\gamma+\omega\right)/2}\\
n_{4}\left(t\right) & = & \frac{\gamma\left(1-e^{-t\gamma_{b}}\right)-\gamma_{b}\left(1-e^{-t\gamma}\right)}{\gamma-\gamma_{b}},\end{eqnarray*}
 where $C_{i},\, C'_{i},\, B_{i}$ are only functions of $J,\,\gamma,\,\gamma_{b}$,
and $C_{3},\, C'_{3},\, B_{1}$ are functions of time as well. Finally
$\omega=\sqrt{\gamma^{2}-4J^{2}}$, resulting in an imaginary eigenvalue
of the Liouvillian for $2\left|J\right|>\gamma$. This in turn shows
up in an oscillating behavior of the populations as a function of
time. In Figure \ref{fig:n3_2exc}, the behavior of population 3 as
a function of time and $\gamma_{b}$ is plotted for the given values
of $J$ and $\gamma$. For large values of $t$, one can observe a
non-monotonic behavior as a function of $\gamma_{b}$ emphasized in
the bottom panel of Figure \ref{fig:n3_2exc}. This behavior can be
qualitatively understood as follows. Consider the behavior of $n_{3}$
as a function of $\gamma_{b}$ for a large fixed time $\tilde{t}$.
Since the effect of $\gamma_{b}$ is that of taking away particles
from site 3, $n_{3}$ first decreases when $\gamma_{b}$ is increased
from zero at fixed $\tilde{t}$. Anyway, if $\gamma_{b}$ is further
increased, excitations are taken away at a faster rate and transferred
to site 4. This means that at the fixed time $\tilde{t}$ site 4 tends
to get full for large $\gamma_{b}$, thus inhibiting the effect of
$L_{b}$. Population $n_{3}$ then increases with $\gamma_{b}$. When
$\gamma_{b}$ is further increased, site 4 becomes effectively full
and $L_{b}$ is turned off, the population becomes then independent
of $\gamma_{b}$ and $n_{3}$ saturates.

\begin{figure}
\noindent \begin{centering}
\includegraphics[width=7cm]{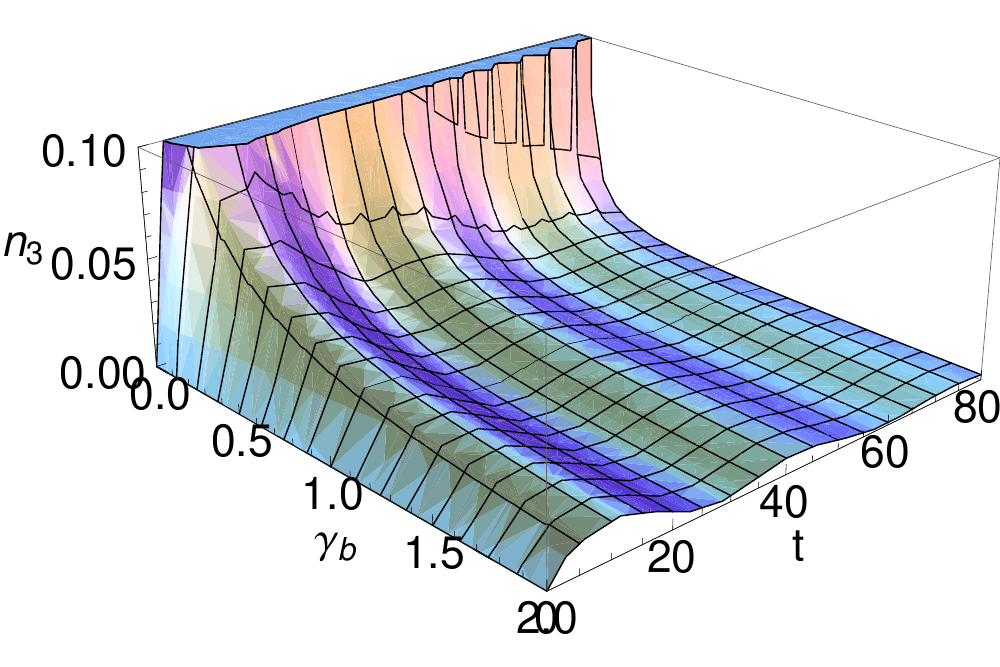} 
\par\end{centering}

\caption{$n_{3}$, as a function of time and $\gamma_{b}$. Initial state has
one excitation at sites 1: $|1,0,0,0\rangle$. Parameters are $J=1,\,\gamma=0.05$.\label{fig:n3_1exc}}

\end{figure}

For the sake of completeness we also consider the solution with one
excitation localized at site 1, i.e.~$|1,0,0,0\rangle$ at time $t=0$.
In this case the time-dependence of the populations is, \begin{eqnarray*}
n_{1}\left(t\right) & = & \frac{e^{-t\gamma/2}}{\omega^{3}}\left[-2J^{2}\omega+\left(\gamma^{2}-2J^{2}\right)\omega\cosh\left(\frac{t\omega}{2}\right)\right.\\
 &  & \left.+\gamma\omega^{2}\sinh\left(\frac{t\omega}{2}\right)\right]\\
n_{2}\left(t\right) & = & \frac{2J^{2}e^{-t\gamma/2}}{\omega^{2}}\left[\cosh\left(\frac{t\omega}{2}\right)-1\right]\\
n_{3}\left(t\right) & = & A_{1}e^{-t\gamma_{b}}+A_{2}e^{-t\gamma}+A_{3}e^{-t\left(\gamma+\omega\right)/2}+A_{4}e^{-t\left(\gamma-\omega\right)/2}\\
n_{4}\left(t\right) & = & 1-\sum_{i=1}^{3}n_{i}\left(t\right)\end{eqnarray*}
 where $A_{i}$ are time independent functions of the parameters.
One can see in Fig.~\ref{fig:n3_1exc} that the non-monotonic behavior
of $n_{3}$ as a function of $\gamma_{b}$ is for this initial condition
absent. As expected, since in the network there are no-excitations
enough to fill the reaction centre, the {}``congestion effect''
is now absent. 

\section{The staircase effect\label{sec:staircase}}

In this section, we explore the situation where excitons are fed into
a quantum network at a given constant rate $\gamma_{\mathrm{in}}$
and are extracted at a rate $\gamma_{\mathrm{out}}$. 

This model can be justified by the fact that some photosynthetic complexes
such as purple bacteria and green-sulfur bacteria \cite{Ganapathy}
live in low-light conditions. The electron-transfer event that occurs
in the reaction center is a process that takes place in the order
of picoseconds. We therefore take the common practice of modeling
the reaction center as an incoherent trap \cite{Mohseni2008}.

\paragraph*{Injection-extraction: $\stackrel{\gamma_{\mathrm{in}}}{\rightsquigarrow}\bullet\stackrel{J}{\leftrightarrow}\bullet\stackrel{\gamma_{\mathrm{out}}}{\rightsquigarrow}$.}

Here, we consider the simplest model for the injection and extraction
of an exciton. The model corresponds to two sites coupled coherently
via the hopping Hamiltonian, $H=\left(J/2\right)\left(\sigma_{1}^{-}\sigma_{2}^{+}+\sigma_{1}^{+}\sigma_{2}^{-}\right)$.
Besides the coherent evolution term, an incoherent injection of excitons
is given by a jump operator $L_{\mathrm{in}}=\sqrt{\gamma_{\mathrm{in}}}\sigma_{1}^{+}$
which injects particles at a rate $\gamma_{\mathrm{in}}$ and a corresponding
incoherent extraction term $L_{\mathrm{out}}=\sqrt{\gamma_{\mathrm{out}}}\sigma_{2}^{-}$.

\begin{figure}
\noindent \begin{centering}
\includegraphics[width=6cm]{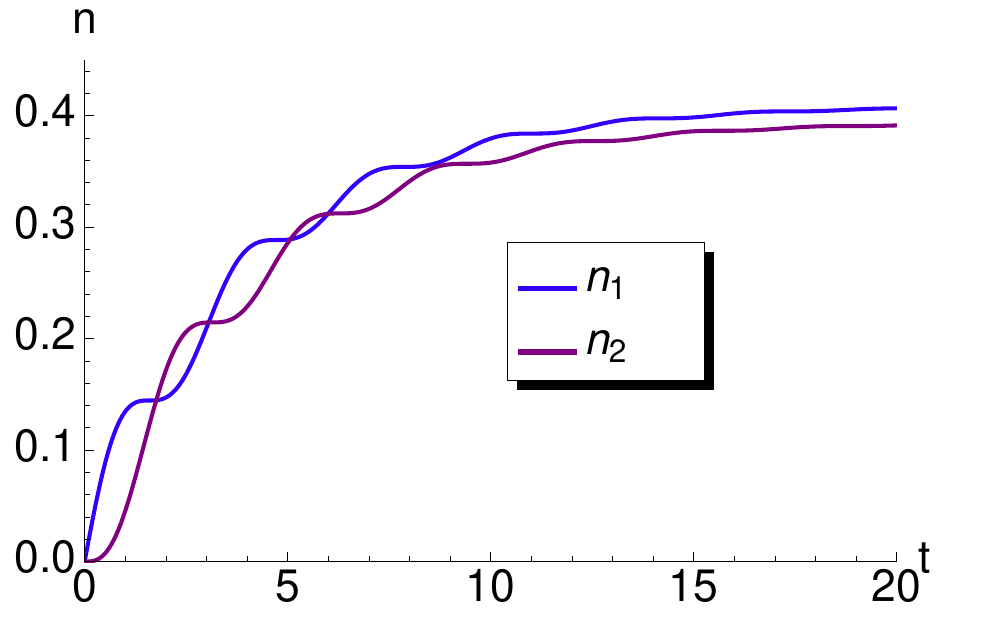} 
\par\end{centering}

\noindent \begin{centering}
\includegraphics[width=5cm]{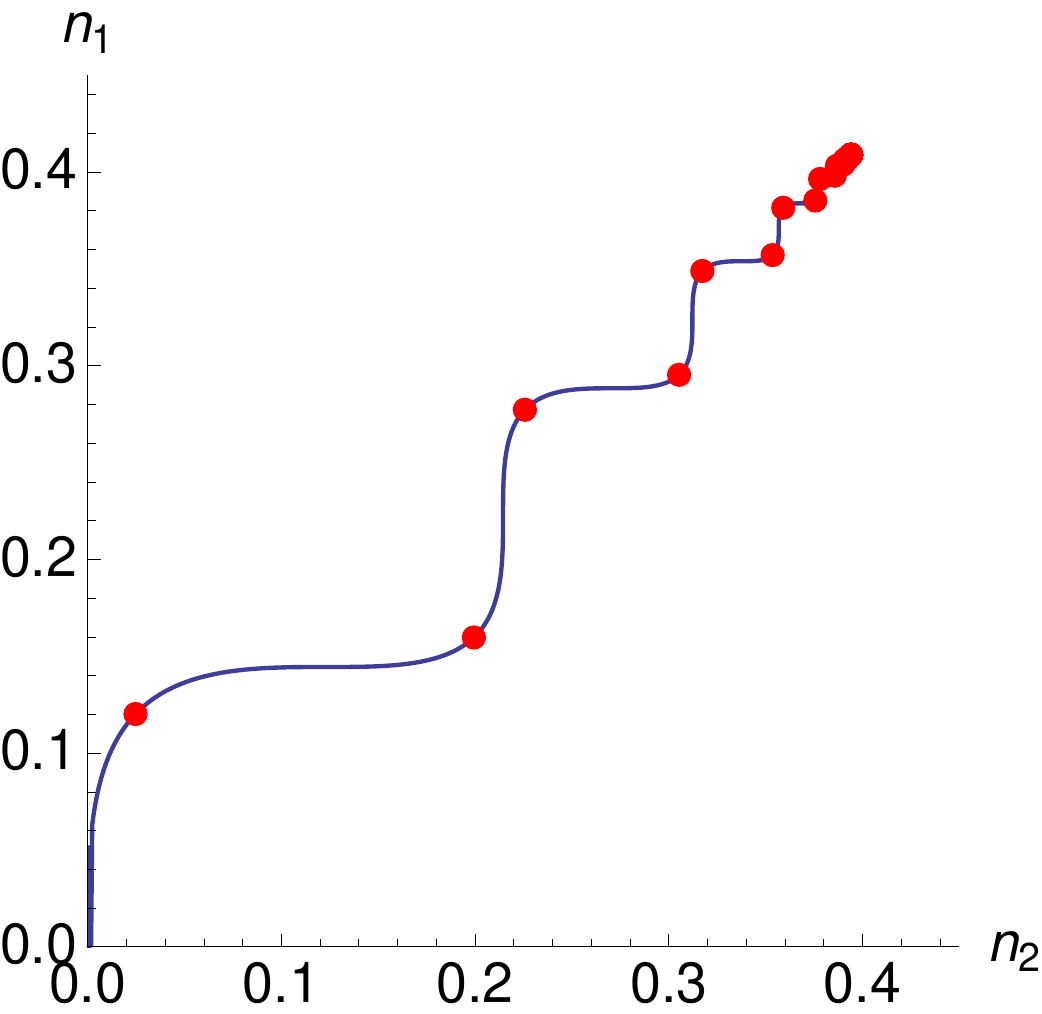} 
\par\end{centering}

\caption{$J=2,$ $\gamma_{\mathrm{in}}=0.2$ (injection), and $\gamma_{\mathrm{out}}=0.3$
(dissipation). Bottom: parametric plot with same parameters. The red
dots correspond to times given by $T_{n}=\left(1/2+n\right)T_{0}$,
$T_{0}=2\pi/\omega,$ ($n=0,1,\ldots$) and the correct frequency
is $\omega=\sqrt{4J^{2}-\left(\gamma_{\mathrm{in}}-\gamma_{\mathrm{out}}\right)^{2}}$.
As explained in the text using general arguments, $T_{0}=O\left(J^{-1}\right)$.
\label{fig:Pump-dissipation-00}}

\end{figure}

The corresponding $16\times16$ Lindblad superoperator matrix can
be diagonalized. A complex eigenvalue with a non-zero imaginary part
gives rise to oscillating behavior in the populations when $\left|\gamma_{\mathrm{in}}-\gamma_{\mathrm{out}}\right|<2\left|J\right|$.

Let us first concentrate on the asymptotic state of the evolution,
$\rho\left(t\rightarrow\infty\right)$. Solving $\mathcal{L}_{\mathrm{tot}}\left(\rho\right)=0,$
one realizes that the asymptotic state is unique and independent of
the initial state. Although this feature is expected in natural physical
systems and follows, for instance, from the detailed balance hypothesis,
it is not necessarily satisfied in our simple toy models (see e.g.~Sec.~\ref{sec:congestion}).

In the standard basis, $\left\{ |1,1\rangle,\,|1,0\rangle,\,|0,1\rangle,\,|0,0\rangle\right\} $,
the explicit expression of the asymptotic state is\begin{multline*}
\rho\left(\infty\right)=\frac{1}{\left(\gamma_{\mathrm{in}}+\gamma_{\mathrm{out}}\right)\left(J^{2}+\gamma_{\mathrm{in}}\gamma_{\mathrm{out}}\right)}\times\\
\left(\begin{array}{cccc}
\frac{J^{2}\gamma_{\mathrm{in}}^{2}}{\left(\gamma_{\mathrm{in}}+\gamma_{\mathrm{out}}\right)} & 0 & 0 & 0\\
0 & \frac{\gamma_{\mathrm{in}}\gamma_{\mathrm{out}}\left(J^{2}+\left(\gamma_{\mathrm{in}}+\gamma_{\mathrm{out}}\right)^{2}\right)}{\left(\gamma_{\mathrm{in}}+\gamma_{\mathrm{out}}\right)} & iJ\gamma_{\mathrm{in}}\gamma_{\mathrm{out}} & 0\\
0 & -iJ\gamma_{\mathrm{in}}\gamma_{\mathrm{out}} & \frac{J^{2}\gamma_{\mathrm{in}}\gamma_{\mathrm{out}}}{\left(\gamma_{\mathrm{in}}+\gamma_{\mathrm{out}}\right)} & 0\\
0 & 0 & 0 & \frac{J^{2}\gamma_{\mathrm{out}}^{2}}{\left(\gamma_{\mathrm{in}}+\gamma_{\mathrm{out}}\right)}\end{array}\right).\end{multline*}
 The only non-vanishing correlations are $\langle\sigma_{1}^{z}\sigma_{2}^{z}\rangle$,
$\langle\sigma_{1}^{z}\rangle$ and $\langle\sigma_{2}^{z}\rangle$.
Thus this state is separable but has non vanishing classical correlations:
$\langle\sigma_{1}^{z}\sigma_{2}^{z}\rangle-\langle\sigma_{1}^{z}\rangle\langle\sigma_{2}^{z}\rangle\neq0$.
Equivalently, the asymptotic state is a classical mixture of states
with definite populations.

Having $\rho\left(\infty\right)$ we can compute the asymptotic populations:\begin{eqnarray}
n_{1}\left(\infty\right) & = & \frac{\gamma_{\mathrm{in}}\left(J^{2}+\gamma_{\mathrm{in}}\gamma_{\mathrm{out}}+\gamma_{\mathrm{out}}^{2}\right)}{\left(\gamma_{\mathrm{in}}+\gamma_{\mathrm{out}}\right)\left(J^{2}+\gamma_{\mathrm{in}}\gamma_{\mathrm{out}}\right)}\label{eq:asymptoticA}\\
n_{2}\left(\infty\right) & = & \frac{\gamma_{\mathrm{in}}J^{2}}{\left(\gamma_{\mathrm{in}}+\gamma_{\mathrm{out}}\right)\left(J^{2}+\gamma_{\mathrm{in}}\gamma_{\mathrm{out}}\right)}.\label{eq:asymptoticB}\end{eqnarray}

A few simple facts can be directly seen from equations (\ref{eq:asymptoticA}),
(\ref{eq:asymptoticB}). First, for small $\gamma_{\mathrm{in}}$
populations deviate by $O\left(\gamma_{\mathrm{in}}\right)$ from
zero; vice versa for $\gamma_{\mathrm{out}}$ small populations deviate
by $O\left(\gamma_{\mathrm{out}}\right)$ from one. Instead, when
$J$ is small excitations get loaded at site 1 but take a long time
to reach site 2 so that $n_{1}=1-O\left(J^{2}\right)$, $n_{2}=O\left(J^{2}\right)$.
Finally, for very large $J$ both populations tend to $n_{1}\simeq n_{2}=\gamma_{\mathrm{in}}/\left(\gamma_{\mathrm{in}}+\gamma_{\mathrm{out}}\right)+O\left(J^{-2}\right)$.

Let us now turn to the dynamics and consider first the most interesting
case namely when the initial state is the empty state $|0,0\rangle$.
A typical result is shown in figure \ref{fig:Pump-dissipation-00}.
An interesting feature clearly emerges: when population $n_{\mathrm{1}}$
increases, $n_{2}$ stays almost constant and vice-versa. Such a feature
is particularly evident in the parametric plot. In the lower panel
of figure \ref{fig:Pump-dissipation-00} we also stressed another
peculiarity of this process: the time needed to increase a given population
when the other is constant (i.e.~the horizontal and vertical steps
between two red dots in Fig.~\ref{fig:Pump-dissipation-00}), is
always the same. We call $T_{0}$ this new, emerging, time-scale.
The description of the entire process then is the following. First
particles are injected at site 1 and population at site 2 stays zero
until a time $T_{0}/2$. Next, for $T_{0}/2<t<3/2T_{0}$ the situation
is reverted and population 2 increases while population 1 remains
constant. The process continues in this fashion until an asymptotic
state is reached. Given the shape of the curve in Fig.~\ref{fig:Pump-dissipation-00}
we refer to this situation as {}``staircase effect''. The emerging
time-scale can be given a physical interpretation considering the
limit when both injection and extraction rates are very small. In
this case the only time-scale of the system is given by the time needed
for the excitations to hop from site 1 to site 2. This time is given
by $T_{0}\approx\Delta E^{-1}=O\left(J^{-1}\right)$. In general,
if we substitute the two sites with an open chain of length $L$,
using the same argument we expect (at least for small $\gamma_{\mathrm{in}}$,
$\gamma_{\mathrm{out}}$) that $T_{0}$ will be the time needed for
the excitations to travel from one side of the chain to the other,
i.e.~$T_{0}\approx L/v$ where $v$ is the velocity of quasiparticles.
Of course this picture can be correct only as long as a quasi-particle
description applies (cf.~Sec.~\ref{sub:application-staircase}).

\begin{figure}
\noindent \begin{centering}
\includegraphics[width=6cm]{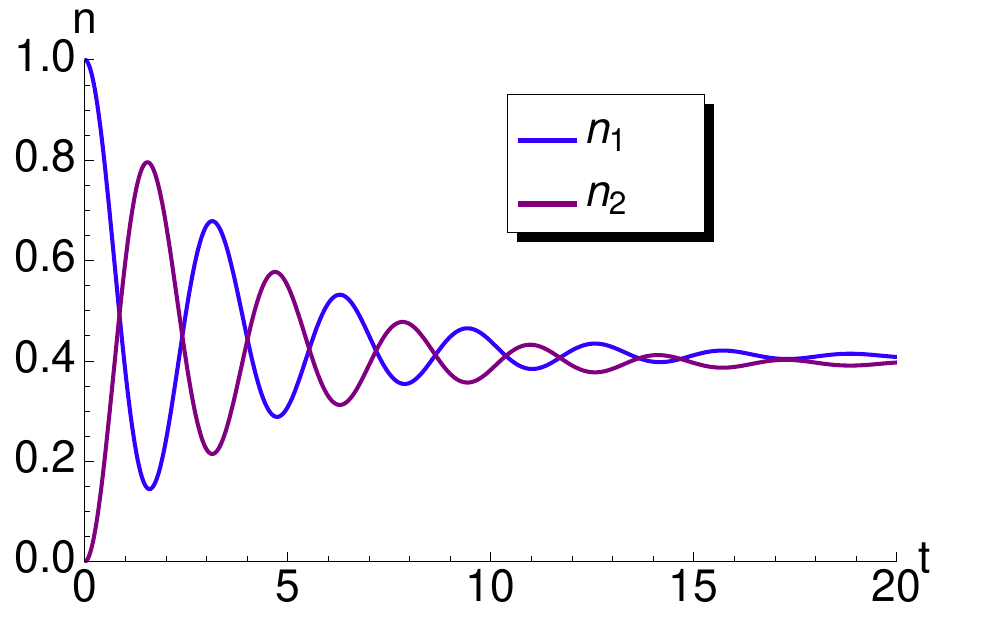} 
\par\end{centering}

\noindent \begin{centering}
\includegraphics[width=5cm]{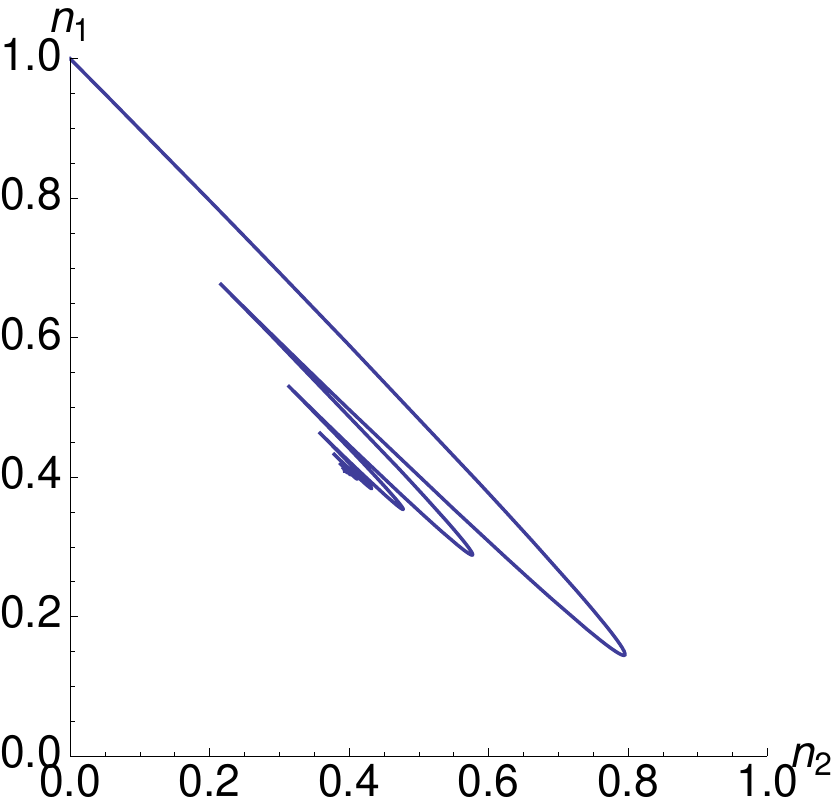} 
\par\end{centering}

\caption{$|1,0\rangle$. Parameters are $J=2$ $\gamma_{\mathrm{in}}=0.2$
(pump) and $\gamma_{\mathrm{out}}=0.3$ (dissipation). Below: parametric
plot, same parameters. \label{fig:Pump-dissipation-10}}

\end{figure}

Let us now consider the injection-extraction dynamics with an initial
state $|1,0\rangle$, i.e.~at time zero the injection site is occupied.
A typical (in the oscillating regime) scenario is shown in Fig.~\ref{fig:Pump-dissipation-10}.
Starting with an initial state $|0,1\rangle$, the situation is almost
identical with $n_{\mathrm{1}}$ and $n_{2}$ interchanged. In fact,
one can show that for initial states with one definite excitation,
populations at any time satisfy the following duality relation\[
n_{1}\left(\gamma_{\mathrm{in}},\gamma_{\mathrm{out}}\right)=1-n_{2}\left(\gamma_{\mathrm{out}},\gamma_{\mathrm{in}}\right).\]
 As previously explained the asymptotic populations do not depend
on the initial populations and are still given by equations (\ref{eq:asymptoticA})
and (\ref{eq:asymptoticB}). The parametric plot in the lower panel
of Fig.~\ref{fig:Pump-dissipation-10} shows that with this initial
condition the staircase effect is not present.

\paragraph*{Three-site injection-extraction: $\stackrel{\gamma_{\mathrm{in}}}{\rightsquigarrow}\bullet\stackrel{J}{\leftrightarrow}\bullet\stackrel{J}{\leftrightarrow}\bullet\stackrel{\gamma_{\mathrm{out}}}{\rightsquigarrow}$.}

A slight generalization of the above idea is given by a three site
chain with injection on the first site and extraction on the third.
For simplicity we consider a uniform chain with equal couplings $J_{12}=J_{23}=J$.
In this case the asymptotic populations are given by\begin{eqnarray*}
n_{1}\left(\infty\right) & = & \frac{\gamma_{\mathrm{in}}\left(J^{2}+\gamma_{\mathrm{in}}\gamma_{\mathrm{out}}+\gamma_{\mathrm{out}}^{2}\right)}{\left(\gamma_{\mathrm{in}}+\gamma_{\mathrm{out}}\right)\left(J^{2}+\gamma_{\mathrm{in}}\gamma_{\mathrm{out}}\right)}\\
n_{2}\left(\infty\right) & = & \frac{\gamma_{\mathrm{in}}\left(J^{2}+\gamma_{\mathrm{out}}^{2}\right)}{\left(\gamma_{\mathrm{in}}+\gamma_{\mathrm{out}}\right)\left(J^{2}+\gamma_{\mathrm{in}}\gamma_{\mathrm{out}}\right)}\\
n_{3}\left(\infty\right) & = & \frac{\gamma_{\mathrm{in}}J^{2}}{\left(\gamma_{\mathrm{in}}+\gamma_{\mathrm{out}}\right)\left(J^{2}+\gamma_{\mathrm{in}}\gamma_{\mathrm{out}}\right)}.\end{eqnarray*}
Note that populations $n_{1}$ and $n_{3}$ are the same as $n_{1}$,
$n_{2}$ in the previous, two-site case. Starting from the totally
empty state, the asymptotic state is reached in a similar manner as
in the two-site case. In particular, the parametric plot of the injection
and extraction sites $\left(n_{1}\left(t\right),n_{3}\left(t\right)\right)$
displays a staircase shape exactly as in the two site case. As we
will show in Sec.~\ref{sub:application-staircase}, this feature
survives even in a longer chain, and is to some extent resistant to
small static diagonal disorder and dephasing.

\section{Asymptotic unitarity\label{sec:Asymptotic-unitarity}}

Another effect we want to study is the possibility that a coherent
dynamics (or sub-dynamics) may emerge out of a dissipative or partly
incoherent dynamics. To make things more clear let us immediately
discuss the simplest example showing this feature.

\paragraph*{Hopping and transfer: $\bullet\stackrel{J}{\leftrightarrow}\bullet\stackrel{\gamma}{\rightsquigarrow}\bullet$.}

The model consists of three sites (qubits). On the first two sites
acts a coherent hopping of the form $H=\left(J/2\right)\left(\sigma_{1}^{-}\sigma_{2}^{+}+\sigma_{1}^{+}\sigma_{2}^{-}\right)$.
On top of that, particles are transferred irreversibly from site 2
to site 3 via a jump operator given by $L=\sqrt{\gamma}\sigma_{2}^{-}\sigma_{3}^{+}$.
It is clear that, if a particle sits at site 3 the incoherent part
of the dynamics is not effective, that is $\mathcal{L}_{L}\left[\rho_{12}\otimes|1\rangle\langle1|\right]=0$.
If we start with an initial state $|1,1,0\rangle$ with sites 1 and
2 occupied and site 3 empty, for effect of the dynamics, site 3 will
get populated at a rate $\gamma$, and on the first two sites there
will remain one particle coherently hopping back and forth. By this
we mean that for a sufficiently large time the evolved state will
be similar to a coherent evolution: $\rho\left(t\right)=e^{t\mathcal{L}_{\mathrm{tot}}}\left[\rho\right]\simeq e^{-itH}\tilde{\rho}e^{itH}=:\tilde{\rho}\left(t\right)$.
For what concerns the state $\tilde{\rho}$ we only know that it will
contain one particle; it can be obtained by evolving back unitarily
$\rho\left(t\right)$, i.e.\[
\tilde{\rho}=\lim_{t\rightarrow\infty}e^{itH}\rho\left(t\right)e^{-itH}.\]
 Indeed, if the dynamics becomes unitary, the above limit is well
defined. Notice that $\tilde{\rho}$ is nothing but the stationary
solution of the original master equation in the interaction picture
associated with $H$. The same reasoning can be done for the subsystem
consisting on sites 1 and 2, i.e.~we can define $\tilde{\rho}_{1,2}$
by evolving back unitarily $\rho_{1,2}\left(t\right)$. Since $H$
does not act on site 3 we have $\tilde{\rho}_{1,2}=\tr_{3}\tilde{\rho}$.
An explicit computation confirms that $\tilde{\rho}=\tilde{\rho}_{1,2}\otimes|1\rangle\langle1|$,
i.e.~in the equivalent, unitary dynamics, one particle sits at site
3. The explicit form of $\tilde{\rho}_{1,2}$ in the standard basis
is\[
\tilde{\rho}_{1,2}=\frac{1}{2\left(J^{2}+\gamma^{2}\right)}\left(\begin{array}{cccc}
0 & 0 & 0 & 0\\
0 & J^{2}+2\gamma^{2} & -iJ\gamma & 0\\
0 & iJ\gamma & J^{2} & 0\\
0 & 0 & 0 & 0\end{array}\right).\]
 This state is a quantum superposition of one-particle states with
$n_{1}=1/2+\gamma^{2}/2\left(J^{2}+\gamma^{2}\right)$ and $n_{2}=1/2-\gamma^{2}/2\left(J^{2}+\gamma^{2}\right)$.

What are the possible indicators of asymptotic unitarity? Since the
purity is constant under unitary evolution, one possibility is to
look at the purity of the total system or of some part of it. The
time-derivative of such a quantity will then be close to zero, for
approximate unitary evolution. Since for Lindbladian evolution the
purity derivative is $\partial_{t}\tr\left(\rho^{2}\right)=2\tr\left[\rho\left(t\right)\mathcal{L}_{\mathrm{tot}}\left(\rho\right)\right]$,
this definition has the advantage of being numerically stable. In
our toy model we have\begin{multline*}
\tr\left\{ \left[\rho_{1,2}\left(t\right)\right]^{2}\right\} =\tr\left[\rho^{2}\right]=1-\frac{J^{2}}{2\left(J^{2}+\gamma^{2}\right)}+\\
\frac{-2e^{-\gamma t}\left(J^{2}+\gamma^{2}+\gamma^{2}\cos\left(Jt\right)\right)+e^{-2\gamma t}\left(3J^{2}+4\gamma^{2}\right)}{2\left(J^{2}+\gamma^{2}\right)}.\end{multline*}

Unfortunately the purity tends to a constant whenever the solution
tends to a constant, as happens, for instance, along the natural process
reaching the asymptotic state. In other words, the smallness of the
purity derivative is a sufficient but not necessary condition for
asymptotic unitarity.

Another possibility is to measure some distance between the actual
state and the one obtained with unitary evolution: $\left\Vert \rho\left(t\right)-\tilde{\rho}\left(t\right)\right\Vert $.
Once again, we might as well restrict to a particular subsystem. Using
the operator norm the result for our toy-model is particularly simple
and illuminating\[
\left\Vert \rho\left(t\right)-\tilde{\rho}\left(t\right)\right\Vert =\left\Vert \rho_{1,2}\left(t\right)-\tilde{\rho}_{1,2}\left(t\right)\right\Vert =e^{-\gamma t}.\]
 This confirms our initial intuition: the dynamics becomes unitary
at a rate $\gamma$. This approach has a very clear meaning but has
the disadvantage of being computationally demanding as it requires
the computation of a matrix norm and the evaluation of $\tilde{\rho}\left(t\right)$.
A simpler alternative is the following.

\begin{figure}
\noindent \begin{centering}
\includegraphics[width=6cm]{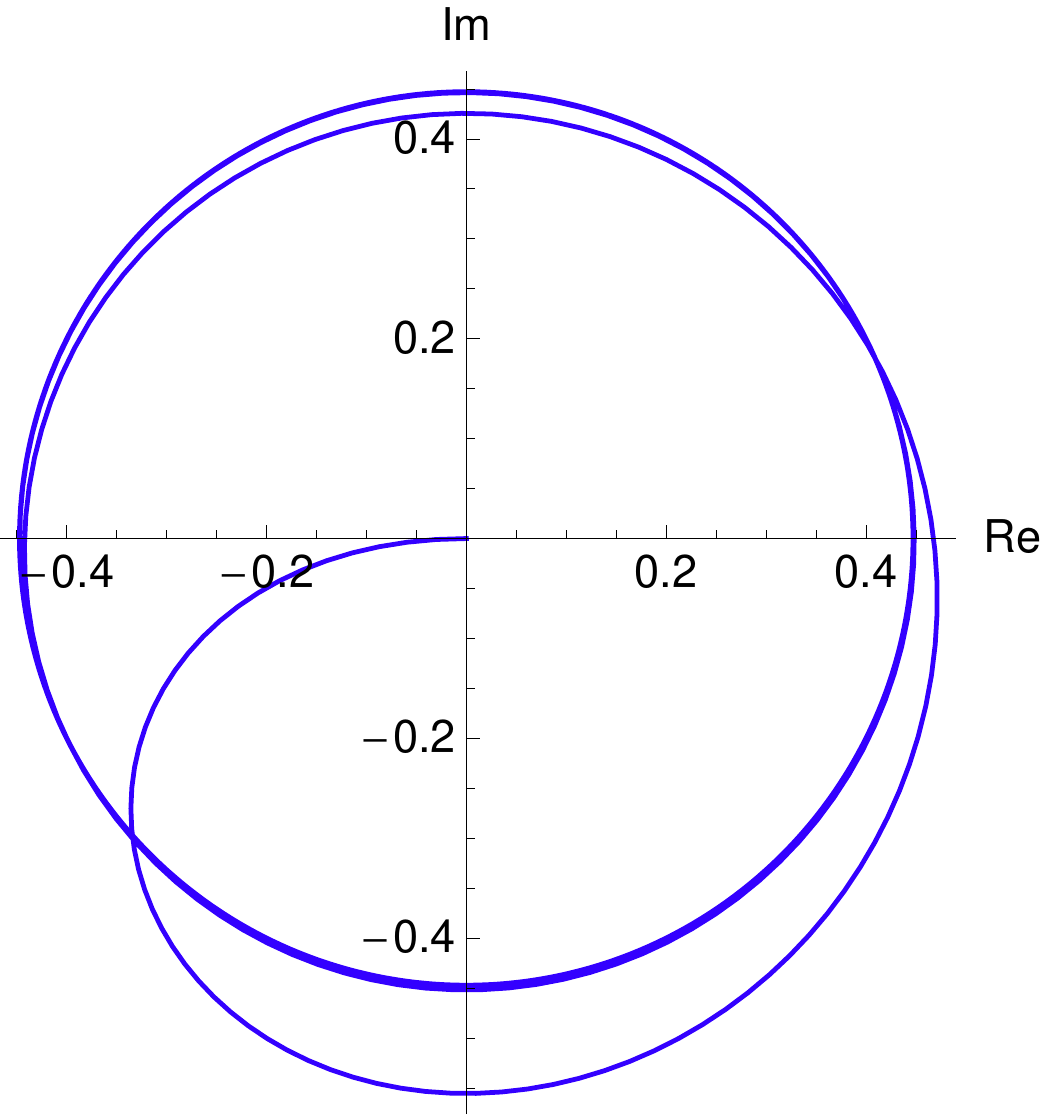} 
\par\end{centering}

\caption{$\langle\psi^{-}|\rho_{1,2}\left(t\right)|\psi^{+}\rangle$ for the
model considered in the text. Parameters are $J=2,\,\gamma=1$. \label{fig:re-im-param}}

\end{figure}

Consider the spectral representation of the Hamiltonian $H=\sum_{n}E_{n}|n\rangle\langle n|$.
If the total evolution becomes similar to a unitary evolution, the
matrix elements of the density matrix in the eigenbasis $|n\rangle$
evolve in time like phases:\[
\langle n|\rho\left(t\right)|m\rangle\simeq\langle n|\tilde{\rho}\left(t\right)|m\rangle=e^{-it\left(E_{n}-E_{m}\right)}\langle n|\tilde{\rho}|m\rangle.\]
 In our model the eigenbasis of the two-site Hamiltonian is $\{|0,0\rangle,\,|1,1\rangle,\,|\psi^{\pm}\rangle=\left(|1,0\rangle\pm|0,1\rangle\right)/\sqrt{2}\}$.
For instance, one can show that \[
\langle\psi^{-}|\rho_{1,2}\left(t\right)|\psi^{+}\rangle=\frac{\gamma}{iJ+\gamma}\left(e^{-t\gamma}-\cos\left(Jt\right)-i\sin\left(Jt\right)\right).\]
 Pictorially the parametric plot of the real and imaginary part of
this matrix element folds on a circle (of radius $\gamma/\sqrt{J^{2}+\gamma^{2}}$)
after a time $\gamma^{-1}$ (see Fig.~\ref{fig:re-im-param}).

This method to mark the appearance of asymptotic unitarity, as well
as the study of the distance $\left\Vert \rho\left(t\right)-\tilde{\rho}\left(t\right)\right\Vert $,
has a major advantage with respect to those based on $\dot{\rho}\left(t\right)$.
Namely it allows to discriminate between approximate unitary evolution
and the usual reach of an asymptotic state for which $\dot{\rho}=0$.

We would like to end this section by stressing the (almost obvious)
relation of asymptotic unitarity with the quantum-information concept
of noiseless or decoherence-free subspace/system \cite{dfs}. The
quantum networks considered in this paper are of hybrid type, namely
some of the inter-site couplings are coherent i.e., hopping, and other
are incoherent i.e., irreversible transfer described by $L.$ On the
other hand, the dynamics restricted to the range of the projection
$P:=\openone_{12}\otimes|1\rangle\langle1|$ is unitary because, as
noticed in the above, $\mathcal{L}_{L}(P\rho P)=0.$ This means that
the range of $P$ is indeed a decoherence-free subspace. Now the dynamics
is such that, for appropriate initial conditions $\lim_{t\to\infty}n_{3}(t)=1$
or equivalently $\lim_{t\to\infty}\|P\rho(t)P-\rho(t)\|=0$ . This
means that the asymptotic state belongs to the range of $P,$ which
in turn implies the unitary nature of the long-time dynamics.

\section{Applications to LH1-RC complexes\label{sec:Applications-to-LH1-RC}}

In this section we want to check if and how the effects studied so
far can survive in more realistic networks. Specifically, we will
consider models which can be relevant for the description of energy
transfer in photosynthetic systems.

\begin{figure}
\noindent \begin{centering}
\includegraphics[width=7cm]{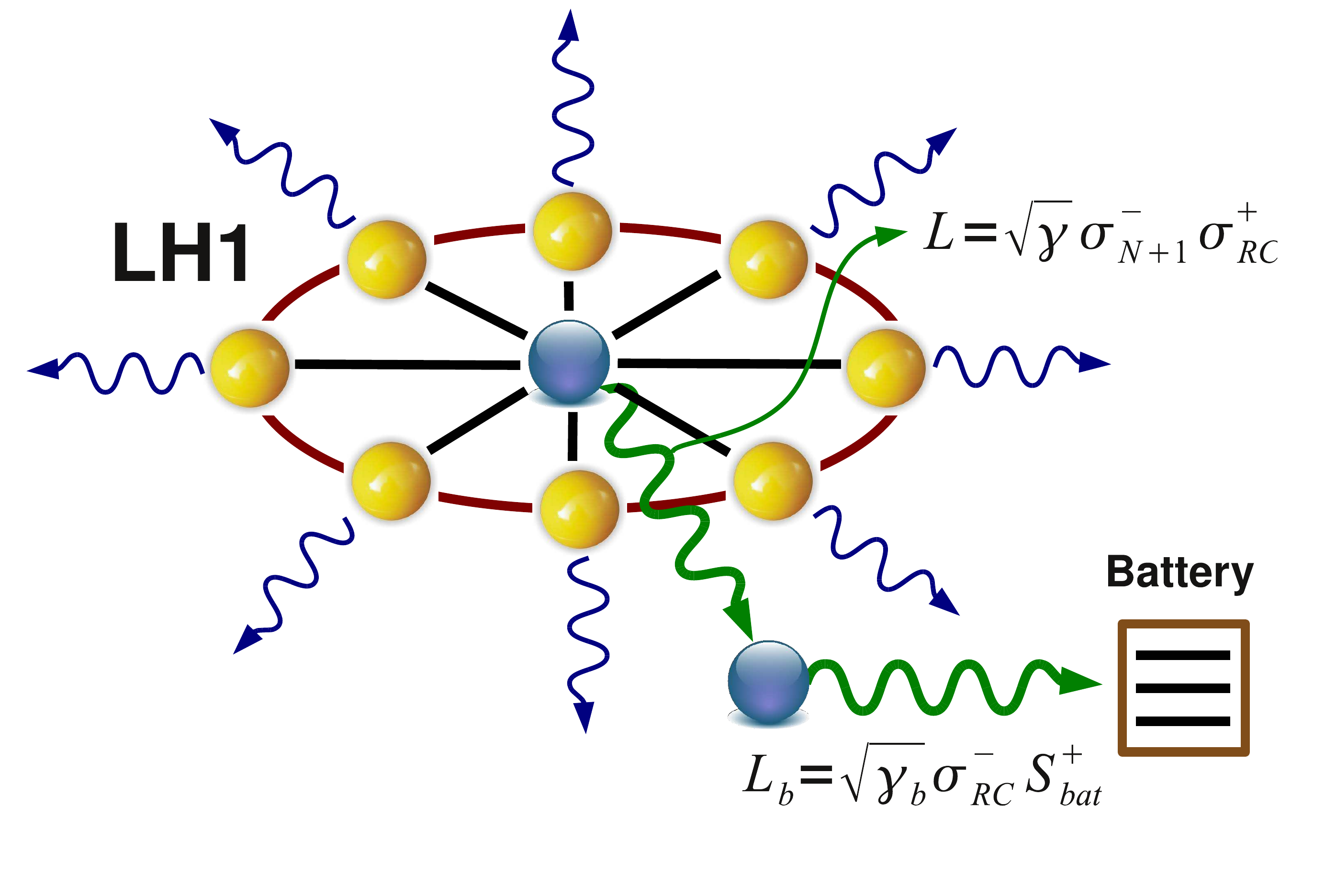} 
\par\end{centering}

\caption{$N$ particles on the ring interact via dimerized nearest neighbor
hopping constants $t_{i,i+1}=t\left(1+\delta\left(-1\right)^{i}\right)$.
Each of these particle can hop to the central site ($N+1$th) with
hopping constant $J$. The central site transfers excitations incoherently
to the reaction center via $L=\sqrt{\gamma}\sigma_{N+1}^{-}\sigma_{RC}^{+}$.
The reaction center itself is connected to a $(2s+1)-$ dimensional
{}``battery'' via $L_{b}=\sqrt{\gamma_{b}}\sigma_{RC}^{-}S_{\mathrm{bat}}^{+}$.
The effect of external degrees of freedom is schematized by incoherent
dissipation and dephasing processes (thin, blue, wavy lines). In actual
LH1 complexes the sites on the ring are bacteriochlorophylls, and
$N=32$. \label{fig:Model-for-LH-I}}

\end{figure}

\subsection{Congestion effect}

Our motivation for the study of the {}``congestion effect'' originated
from a careful analysis of the structure of the reaction center in
LH1-RC complexes. In most photosynthetic bacteria, photons are captured
by light-harvesting antennae where a particle-hole exciton is created
and carried to the reaction center (RC) where eventually a redox reaction
takes place \cite{BlankenshipBook}. In the LH1-RC complexes present
in purple bacteria %
\footnote{Purple bacteria are protobacteria which implement photosynthesis without
producing oxygen.%
} the light harvesting complex and the RC form a compact core unit.
Typical transfer times of excitations to the RC are of the order of
hundreds of picoseconds. A cartoon picture of the LH1-RC complex is
shown in figure \ref{fig:Model-for-LH-I}. Yellow spheres represents
the bacteriochlorophylls forming LH1. In the purple bacterium \textit{Rodobacter
sphaeroides}, there are 32 bacteriochlorophylls (BChl) displaced on
a ring surrounding the reaction center. In figure \ref{fig:Model-for-LH-I}
we display a possible structure for the RC. Instead of treating the
RC as a simple two-level system, as typically done in the literature,
we replace the RC with a structure containing two qubits and a $d$-level
system which we call a {}``battery''. In purple bacteria this structure
has to be imagined sitting at the center of the ring. The first of
these qubits (the $N+1$th) interacts via coherent dipole-dipole hopping
with the BChls of the ring. Excitations are then transferred at a
rate $\gamma$ to what we call reaction center. In turn, the RC itself
is connected to larger $d$-level system ($d=3$ in our simulations)
via irreversible transfer at a rate $\gamma_{b}$. It is the interplay
between the two timescales $\gamma^{-1}$ and $\gamma_{b}^{-1}$,
and their relation to the transfer efficiency, that we want to analyze
here.

The master equation for the whole system is of Lindblad type: $\dot{\rho}=-i\left[H,\rho\right]+\mathcal{L}_{\mathrm{tot}}\left(\rho\right)$.
For what we said so far, the incoherent part is given by $\mathcal{L}_{\mathrm{tot}}=\mathcal{L}_{L}+\mathcal{L}_{L_{b}}+\mathcal{L}_{\mathrm{noise}}$
with $L=\sqrt{\gamma}\sigma_{N+1}^{-}\sigma_{RC}^{+}$ and $L_{b}=\sqrt{\gamma_{b}}\sigma_{RC}^{-}S_{bat}^{+}$.
Dissipation and dephasing effects are taken into account via incoherent
terms acting on the sites of the ring\emph{ $\mathcal{L}_{\mathrm{noise}}=\sum_{j=1}^{N}\mathcal{L}_{L_{j,\mathrm{diss}}}+\mathcal{L}_{L_{j,\mathrm{deph}}}$}
with $L_{j,\mathrm{diss}}=\sqrt{\gamma_{\mathrm{diss}}}\sigma_{j}^{-}$
and $L_{j,\mathrm{deph}}=\sqrt{\gamma_{\mathrm{deph}}}\mathfrak{n}_{j}$.

Regarding the Hamiltonian of the ring degrees of freedom, we referred
to the detailed structure of couplings given in \cite{hu97,hu98}.
The most salient feature emerging from the data of \cite{hu98} is
that the couplings present a dimerized structure: strong coupling
$t_{+}=t\left(1+\delta\right)$ alternate with weak ones $t_{-}=t\left(1-\delta\right)$.
Indeed, instead of using all the couplings $t_{i,j}$ reported \cite{hu98},
almost the same band structure can be obtained using only a nearest
neighbor description with a dimerization of $\delta=0.12$. Some groups
have suggested the possibility that dimerization might favor the transfer
efficiency \cite{yang10}. Our choice of resorting to a dimerized
nearest neighbor hopping structure has the additional advantage of
making the system scalable to different sizes $N$. Hence our choice
for the Hamiltonian is \[
H=\sum_{j=1}^{N}t_{j}\left(\sigma_{j}^{-}\sigma_{j+1}^{+}+\sigma_{j}^{+}\sigma_{j+1}^{-}\right)+J\left(\sigma_{j}^{-}\sigma_{N+1}^{+}+\sigma_{j}^{+}\sigma_{N+1}^{-}\right).\]
 This represents $N$ particles on a ring hopping between neighboring
sites with constants $t_{j}=t\left(1+\delta\left(-1\right)^{j}\right)$
and to a central site $N+1$ with hopping constant $J$. We will also
add static random diagonal noise ($H\rightarrow H+\sum_{j}\epsilon_{j}\mathfrak{n}_{j}$)
to inhibit the possible appearance of decoherence-free subspaces which
can limit the efficiency of transfer \cite{PlenioHuelga2009}.

\begin{figure}
\begin{centering}
\includegraphics[width=4cm]{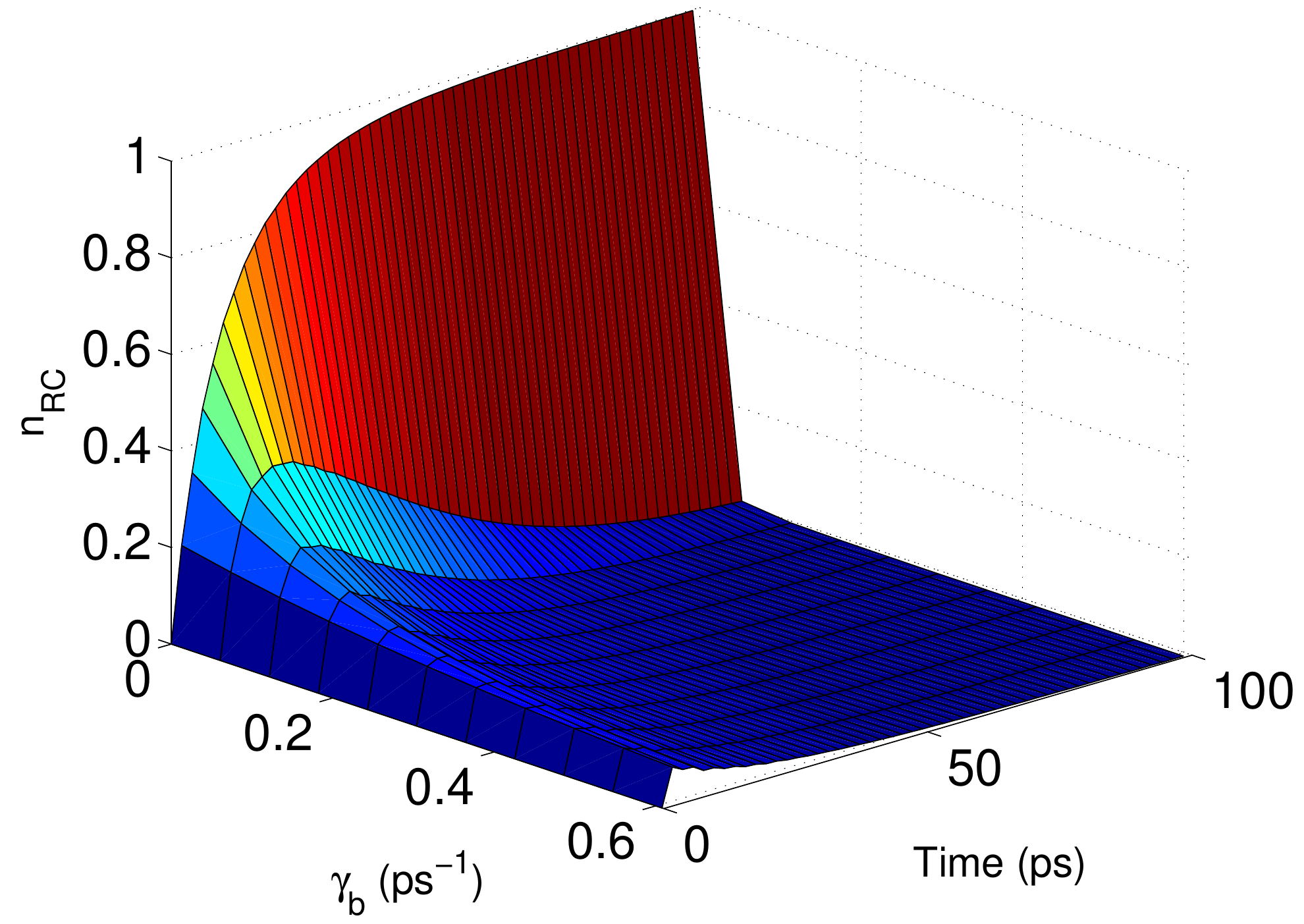}\includegraphics[width=4cm]{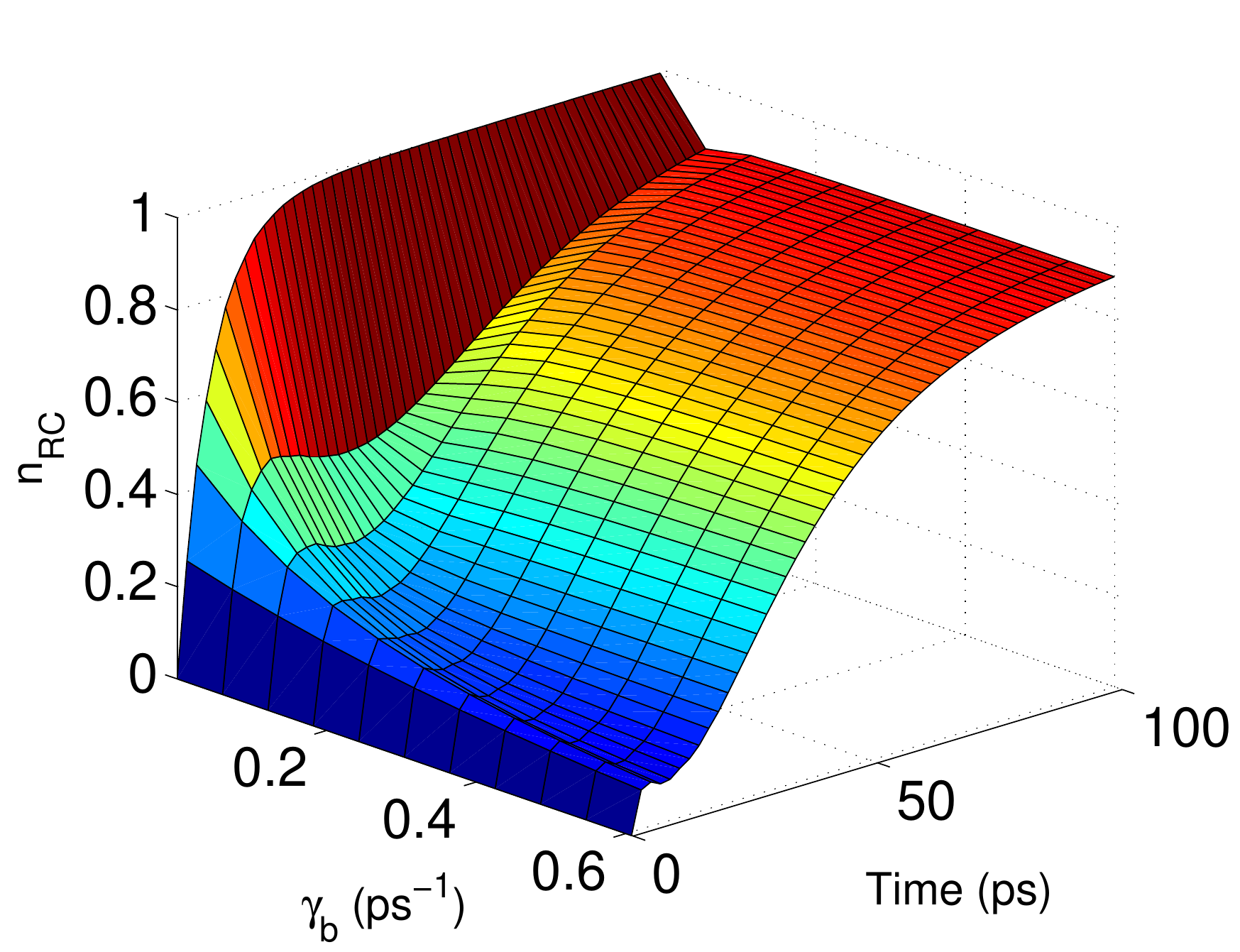} 
\par\end{centering}

\begin{centering}
\includegraphics[width=4cm]{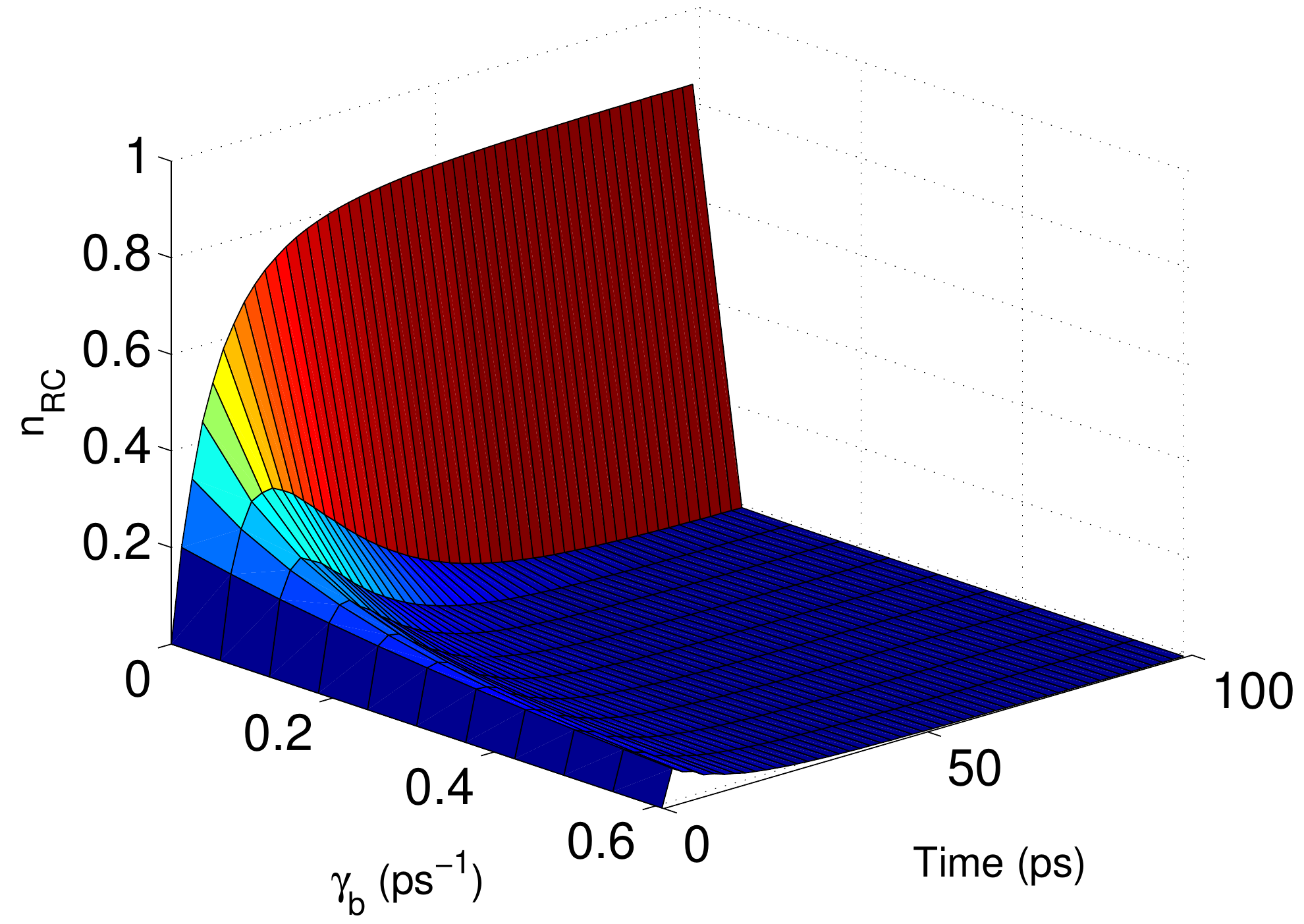}\includegraphics[width=4cm]{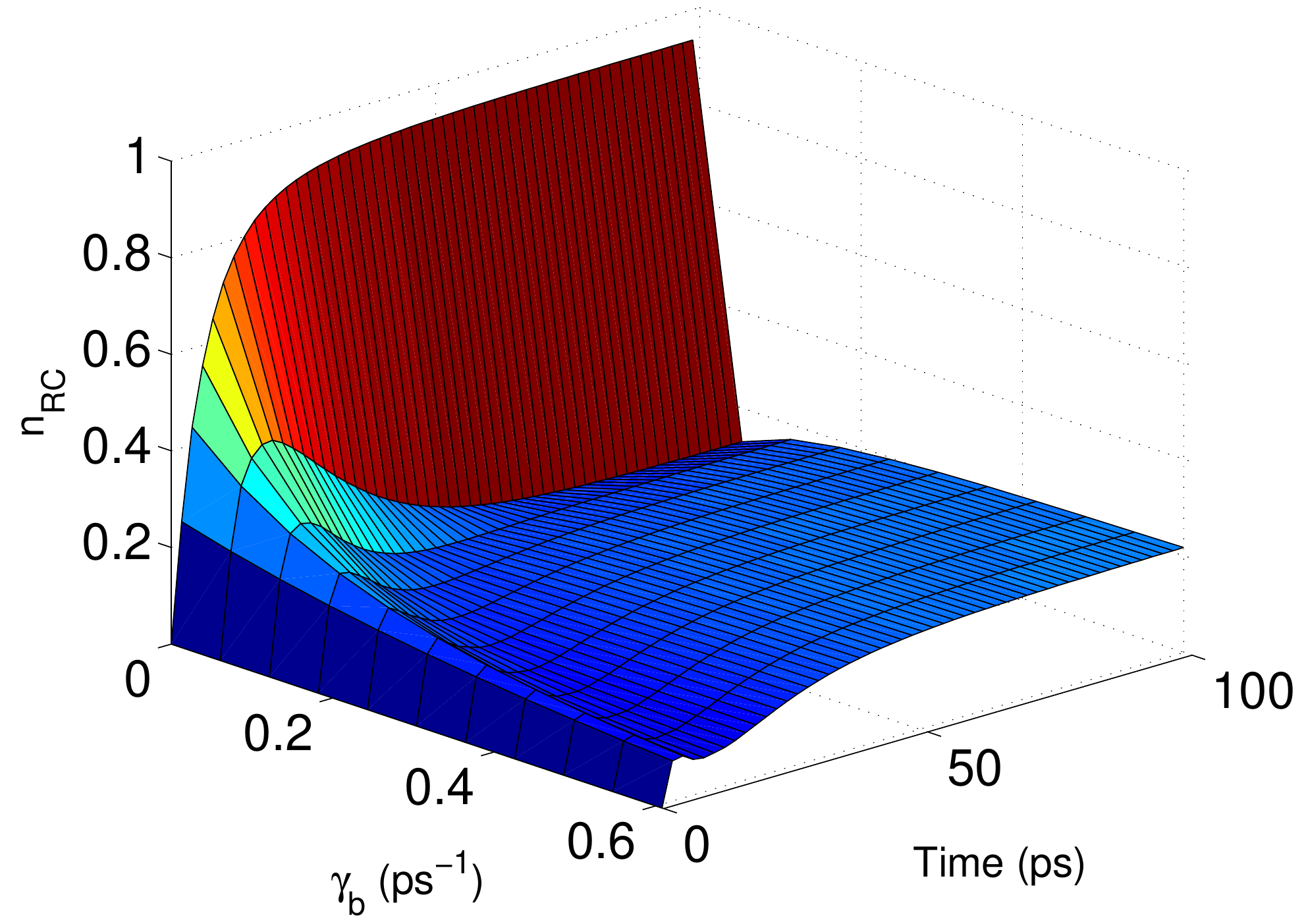} 
\par\end{centering}

\caption{{}``congestion effect'' in light-harvesting complexes. Upper panels:
clean system, no dissipation and dephasing. The ring has $N=4$ sites,
hopping constants are $t=J=1\,(meV)$ and dimerization is $\delta=0.12$.
Diagonal static noise of the form $\epsilon_{n}=t\cos\left(en\right)$
is added. The RC transfer rate is set to $\gamma=0.3\, ps^{-1}$.
Lower panels: same parameters plus dissipation and dephasing $\gamma_{\mathrm{diss}}=\gamma_{\mathrm{deph}}=0.03\, ps^{-1}$.
Left panels: the initial state is a two-particle Dicke state for the
ring, other sites are empty. Right panels: the initial state is a
three-particle Dicke state for the ring, other sites are empty. \label{fig:congestion_LHI}}

\end{figure}

The results of our simulations are shown in Fig.~\ref{fig:congestion_LHI}.
We initialize the system by starting with a pure Dicke state for the
ring while keeping all other sites empty. This means the initial state
is $|\psi_{0}\rangle=\left(\begin{array}{c}
N\\
n\end{array}\right)^{-1/2}\left(\sigma_{\mathrm{tot}}^{+}\right)^{n}|0\rangle$ where $\sigma_{\mathrm{tot}}^{+}=\sum_{j=1}^{N}\sigma_{j}^{+}$ refers
only to the ring sites and $|0\rangle$ is the empty state for the
whole system. The choice of an initial Dicke state is natural for
a series of reasons. First it allows to treat initial states with
general definite particle number $n\le N$. Second, Dicke states are
symmetric under permutation, thus carrying no net momentum. If the
photon's wavelength is larger than the size of the LH1 complex, the
excitations created must be a completely delocalized $k=0$ object.
In any case, since only the $k=0$ component of the ring couples to
the central $N+1$th site, transfer in the antisymmetric channel $k=\pi$,
being a higher order process, is highly suppressed and gives much
lower transfer efficiency \cite{OlayaCastro2008}.

We first performed simulations on a {}``clean'' system, i.e.~with
no dissipation or dephasing present. In Fig.~\ref{fig:congestion_LHI}
we plotted the population of the reaction center (normally called
efficiency $\eta$ in the literature) as a function of time for different
values of $\gamma_{b}$. Looking at the upper panels of Fig.~\ref{fig:congestion_LHI},
the situation is completely analogous to the congestion effect observed
in our simple toy model (see Figures \ref{fig:n3_2exc}, \ref{fig:n3_1exc}).
As long as we start with a number of excitations which can be accommodated
in the battery, they will all flow to the battery for $\gamma_{b}\neq0$
(left panel). When we start with 3 particles in the ring we see again
the appearance of a non-monotonic behavior between $\gamma_{b}$ and
$\gamma$ which shows up as a valley at large times and $\gamma_{b}\lesssim\gamma$
($\gamma_{b}$ smaller than, but of the order of $\gamma$). When
we add additional decoherence in the form of dissipation and dephasing
the situation is only quantitatively changed. The valley due to the
{}``congestion effect'', although less pronounced, is still visible
in the bottom right panel of Fig.~\ref{fig:congestion_LHI}.

\subsection{Asymptotic unitarity}

\begin{figure}
\noindent \begin{centering}
\includegraphics[width=4cm]{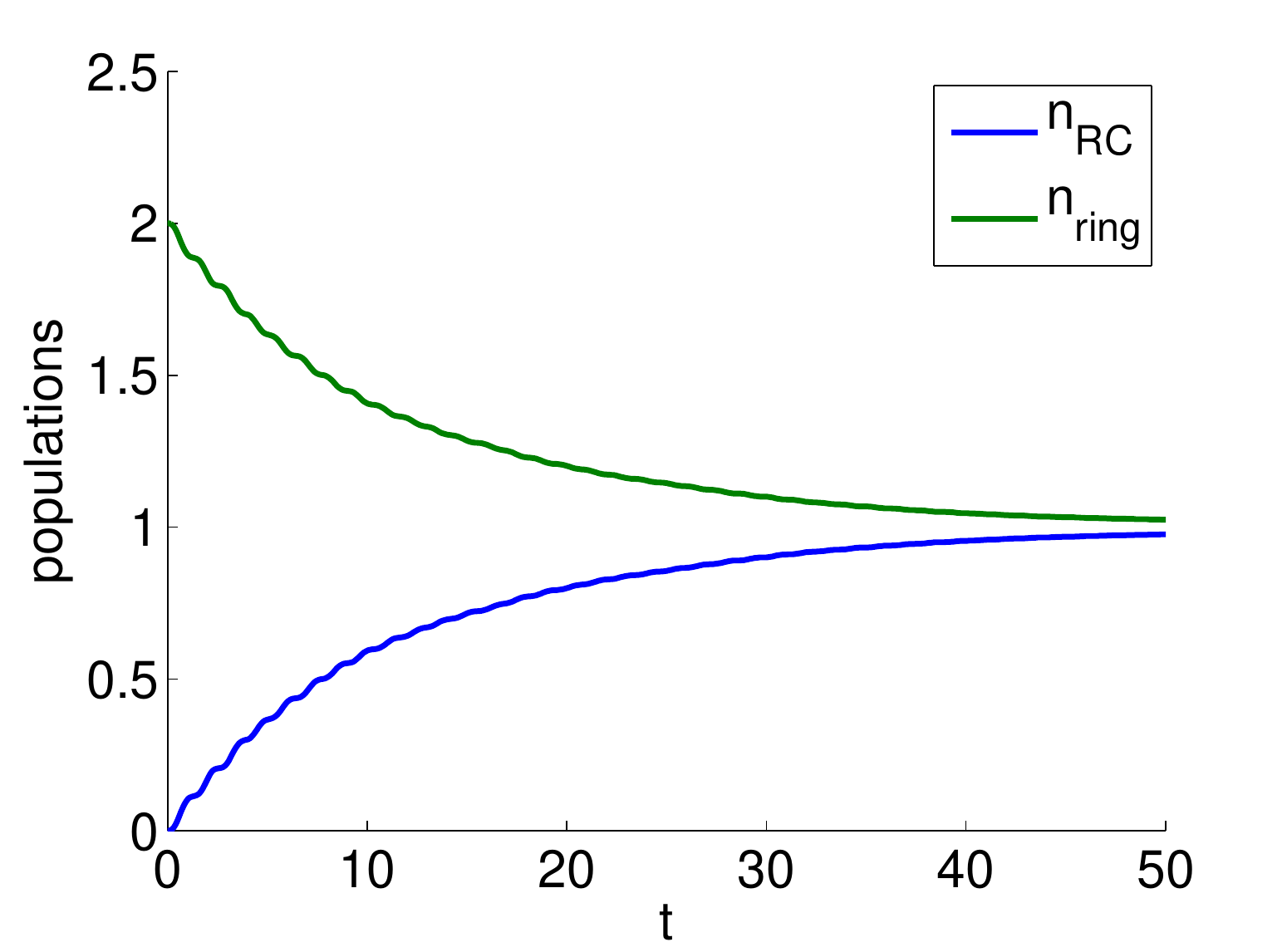}\includegraphics[width=4cm]{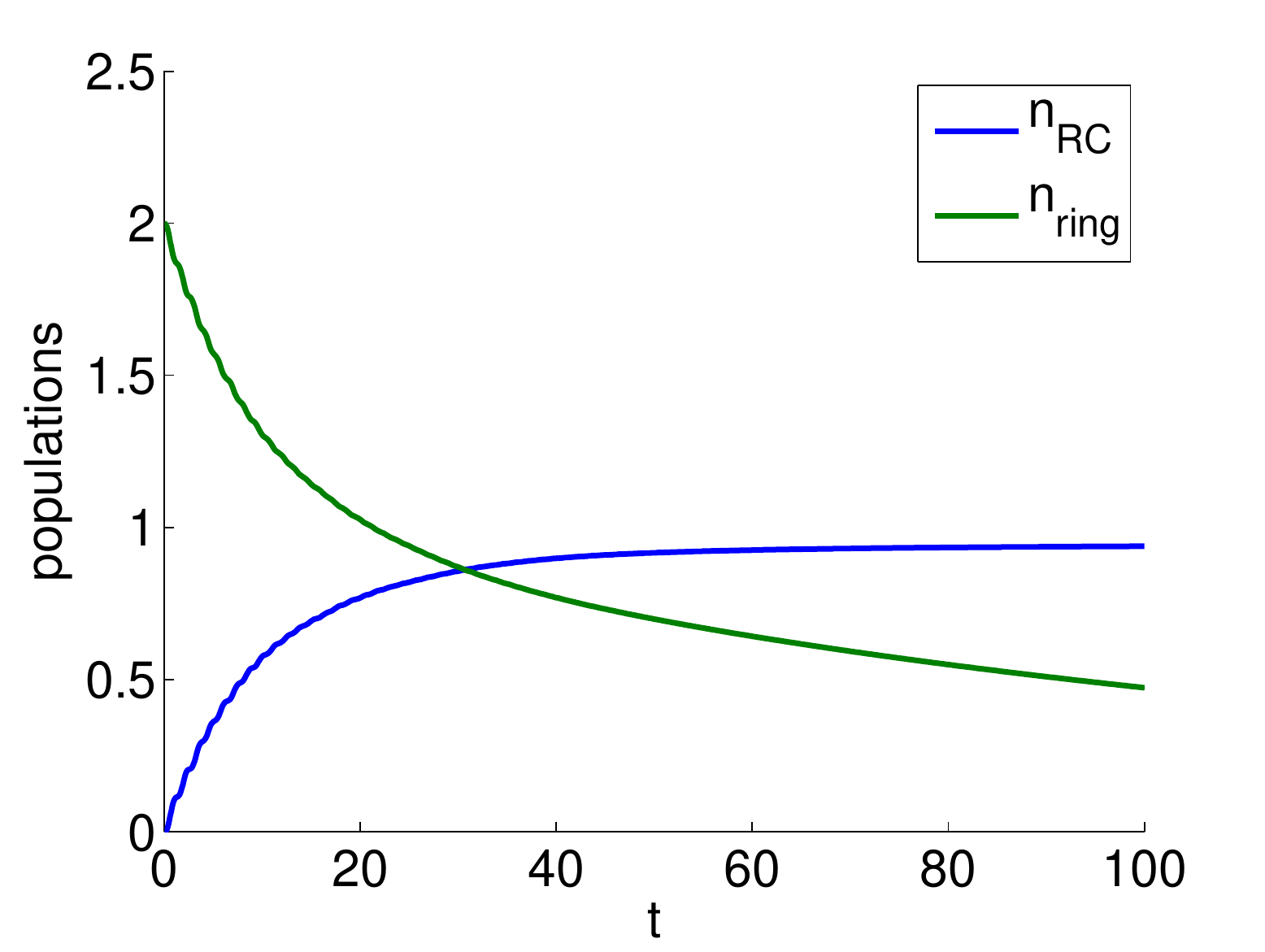} 
\par\end{centering}

\noindent \begin{centering}
\includegraphics[width=4cm]{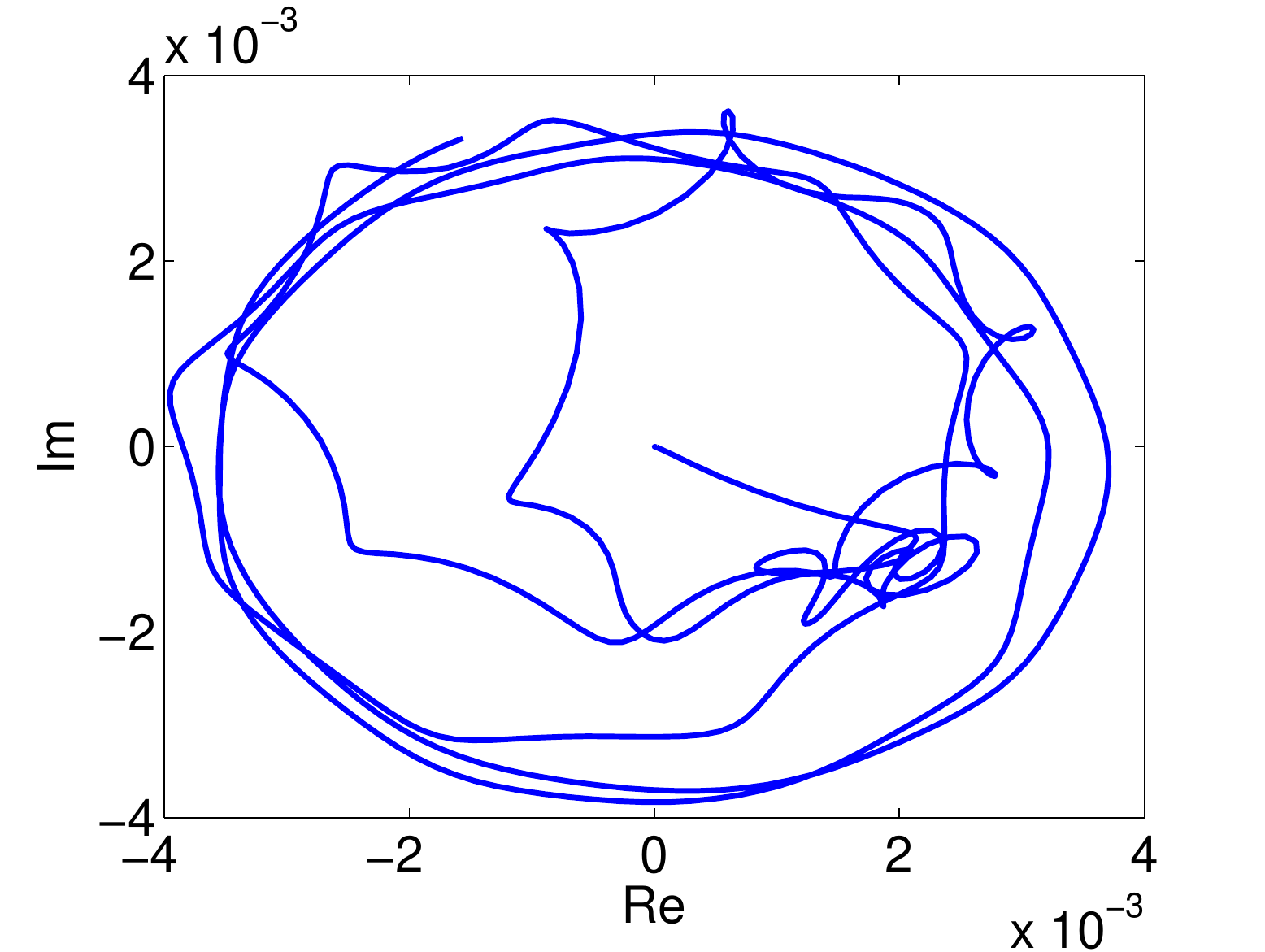}\includegraphics[width=4cm]{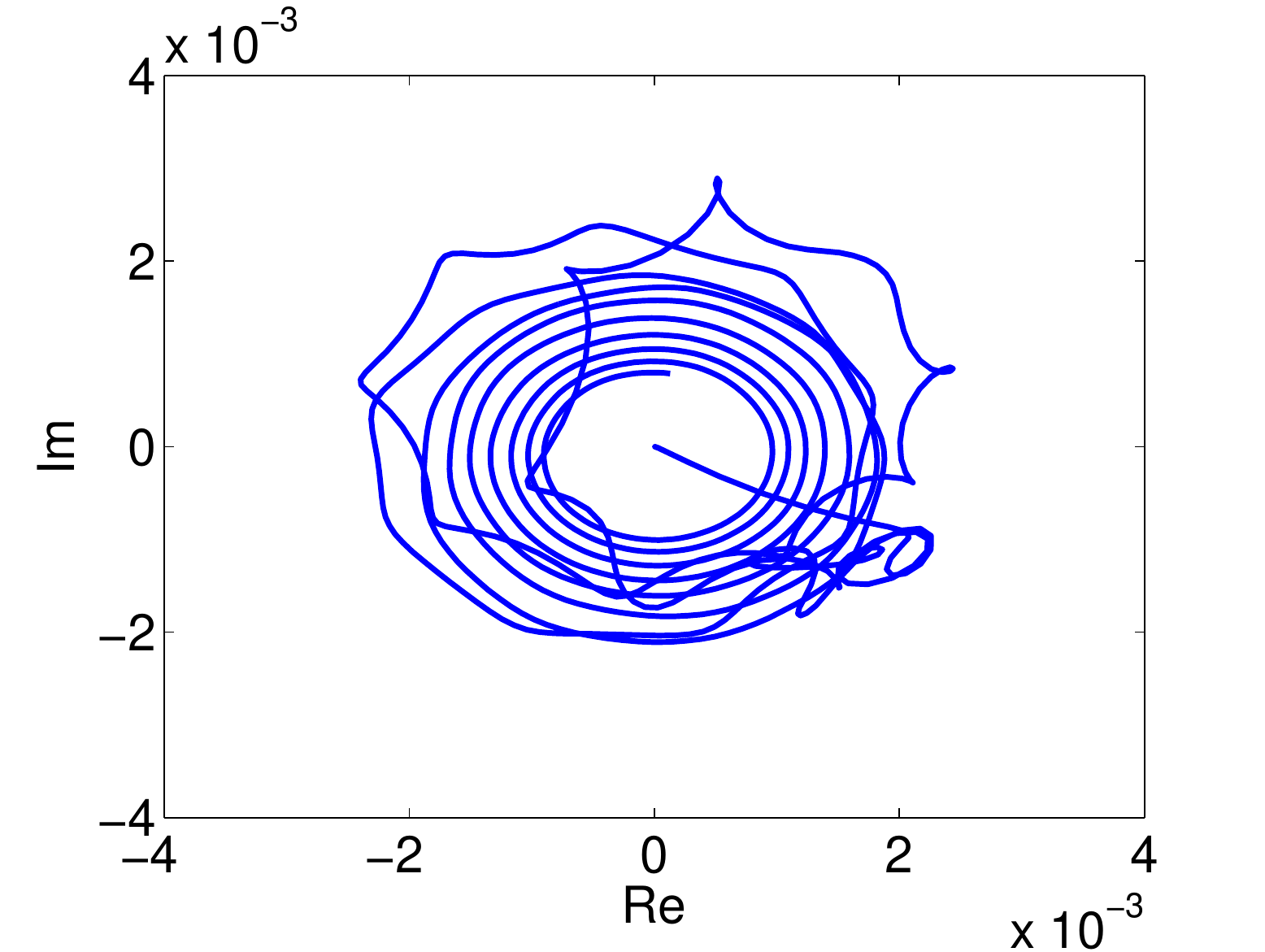} 
\par\end{centering}

\caption{$N+2=6$ sites. On the upper panels we plot the populations of the
ring and of the RC as a function of time (arbitrary units). Bottom
panels: parametric plot of the real and imaginary parts of a matrix
element $\langle\psi_{m}|\langle1|\rho\left(t\right)|\psi_{n}\rangle|1\rangle$
for certain $m,\, n$. $|\psi_{n}\rangle$ are the Hamiltonian eigenstates.
Left panels: the Hamiltonian has $t=J=1$, $\delta=0.12,$ diagonal
static noise $\epsilon_{p}=t\cos\left(ep\right),$ and no dissipation
or dephasing. Excitations are transferred to the RC at a rate $\gamma=0.2$.
Right panels: same parameters, but on the particles of the ring acts
dissipation and dephasing with $\gamma_{\mathrm{deph}}=\gamma_{\mathrm{diss}}=0.01$.
\label{fig:DDFS_dirty}}

\end{figure}

To study asymptotic unitarity the {}``battery'' is an unnecessary
complication. Therefore, we will use the same model of Fig.~\ref{fig:Model-for-LH-I}
without the battery site and the corresponding jump operator. This
leads to a network of $N+2$ qubits where the $N+1$th site is connected
to the RC via irreversible transfer at a rate $\gamma$. As done previously,
we will use an $n$-particle Dicke state as the initial state. Let
us first consider the case where the only incoherent term is the one
transferring particles from the central site to the RC. In this case
the dynamics becomes exactly unitary when the RC is full. Simulations
on a network with $N+2=6$ qubits are shown in Fig.~\ref{fig:DDFS_dirty}.
We also show the effect of dissipation and dephasing, though one order
of magnitude smaller than the RC transfer. For short times the evolution
is the same as for the clean (i.e.~no dissipation and dephasing)
case, however for time of order $\gamma_{\mathrm{diss}}^{-1}$ dissipation
sets in and the parametric plot for a generic matrix element $\langle n|\rho\left(t\right)|m\rangle$,
spirals down to zero (Fig.~\ref{fig:DDFS_dirty} bottom right plot).

The conclusion of this section is as simple as it is intriguing, in
view of potential applications to biological systems. If the time-scale
$\gamma_{\mathrm{diss}}^{-1}$ is large enough, there may exist a
time window $T_{\mathrm{relax}}<t<\gamma_{\mathrm{diss}}^{-1}$ in
which quantum effects are not only visible but the dynamics is effectively
unitary! In our models $T_{\mathrm{relax}}$ is the time needed for
the RC to get filled, and is of the order of $T_{\mathrm{relax}}\sim\gamma^{-1}$.
Even more important is the fact that the incoherent transfer to the
RC must shut down when the RC is full. Considering the LH1-RC complex,
the separation of time-scales does indeed occur. For instance in \cite{OlayaCastro2008,OlayaCastro2010b}
the dissipation is four orders of magnitude smaller than the RC charge-separation
rate. Whether the RC shuts down when it is occupied, although plausible,
is much harder to assess.

\subsection{Staircase effect\label{sub:application-staircase}}

\begin{figure}
\noindent \begin{centering}
\includegraphics[width=7cm]{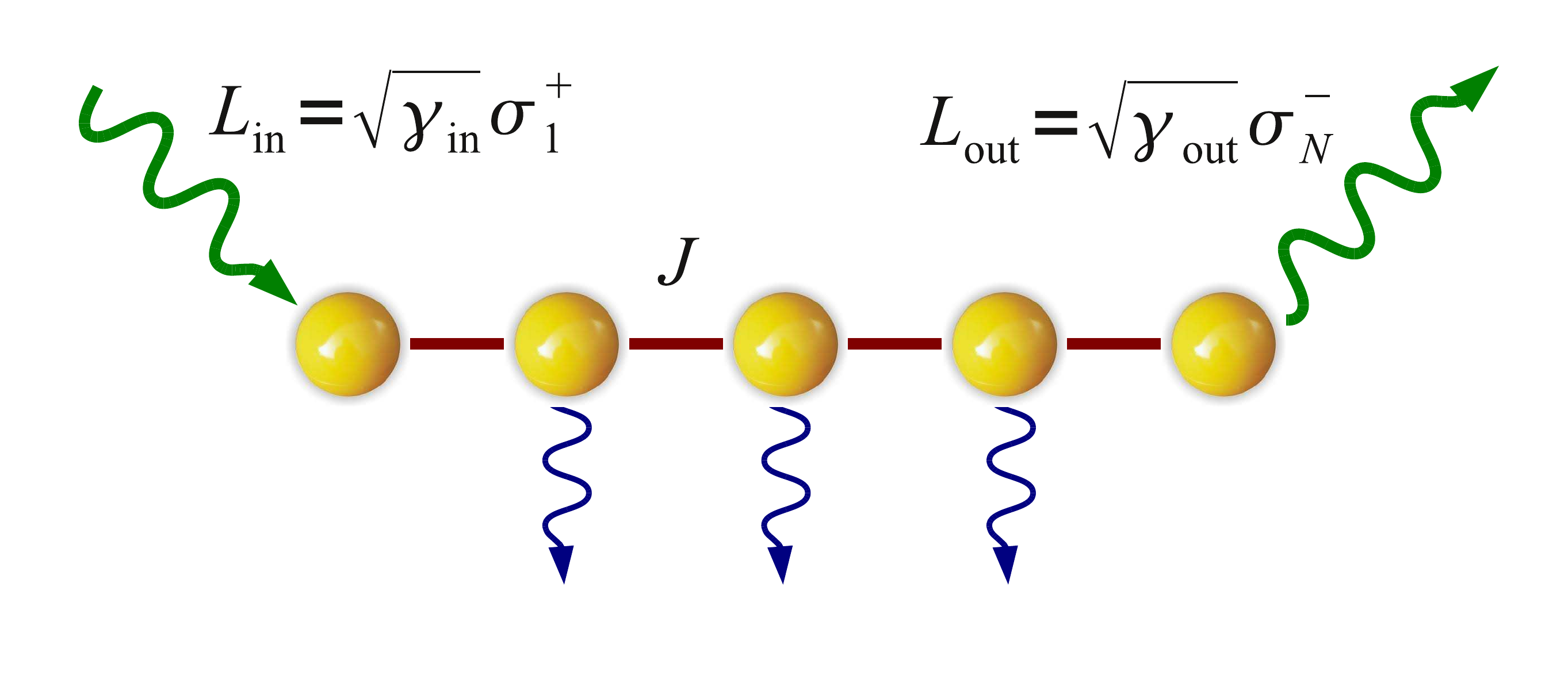} 
\par\end{centering}

\caption{$N$ sites interact via a nearest-neighbor hopping Hamiltonian. Particles
are injected, and respectively expelled incoherently at rates $\gamma_{\mathrm{in}}$,
$\gamma_{\mathrm{out}}$ on the first and last sites . On top of this
basic structure we can add static diagonal disorder and dissipation
as well as dephasing (symbolized by blue wavy arrows) $\mathcal{L}_{\mathrm{noise}}=\sum_{j=2}^{N-1}\mathcal{L}_{L_{j,\mathrm{diss}}}+\mathcal{L}_{L_{j,\mathrm{deph}}}$.
\label{fig:ladder-picture}}

\end{figure}

\begin{figure}
\noindent \begin{centering}
\includegraphics[width=4.5cm]{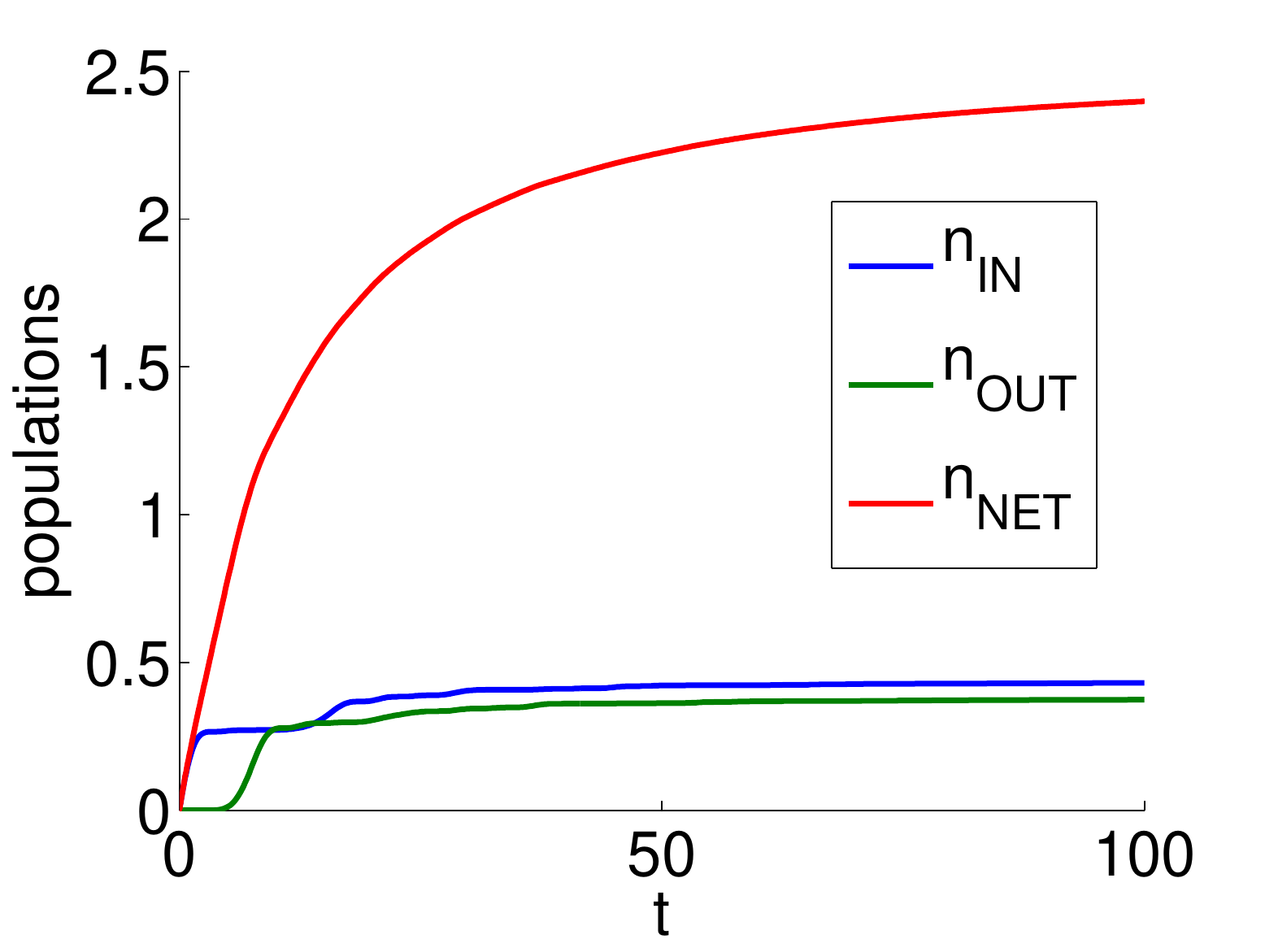}\includegraphics[width=4.5cm]{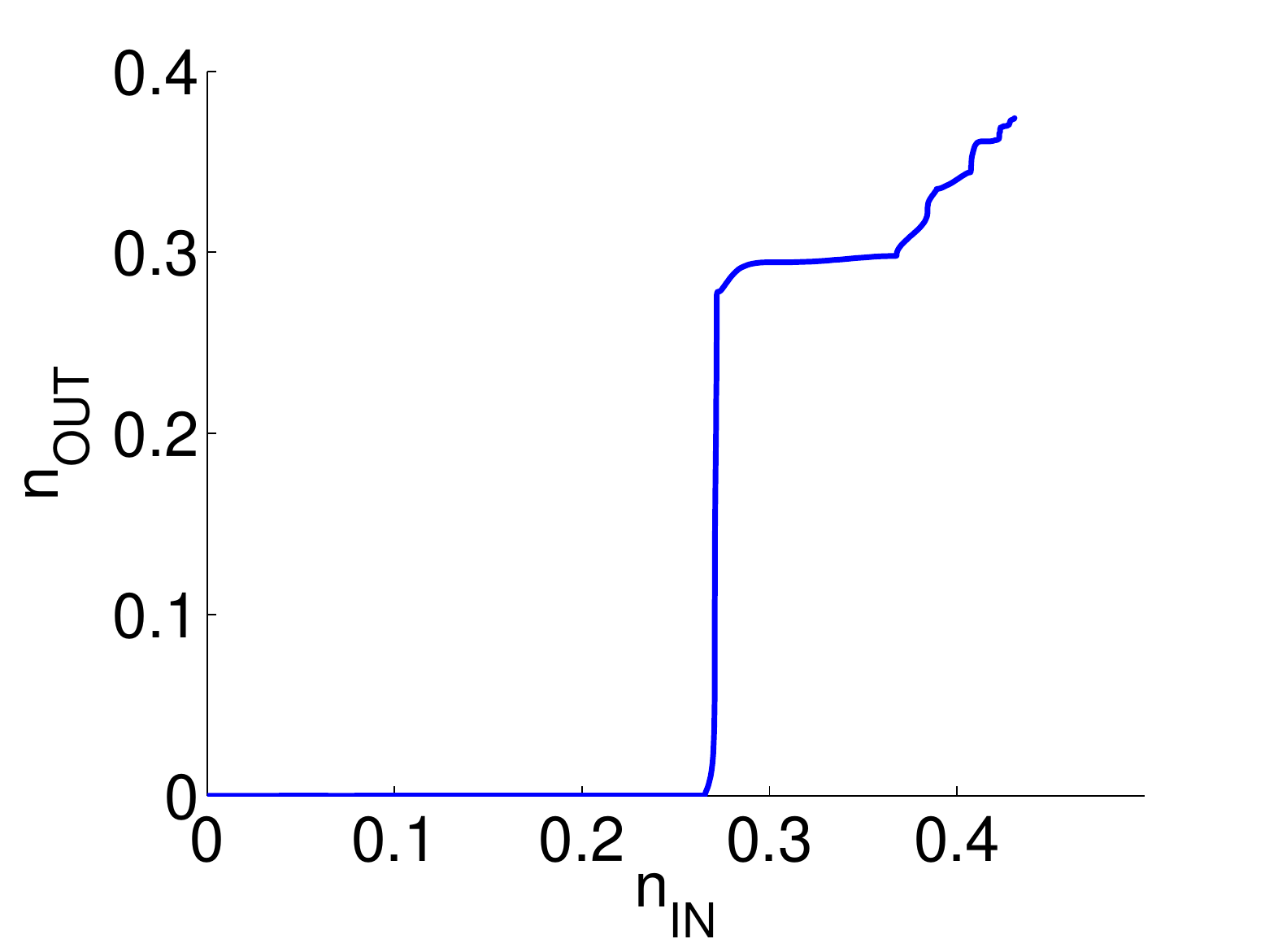} 
\par\end{centering}

\noindent \begin{centering}
\includegraphics[width=4.5cm]{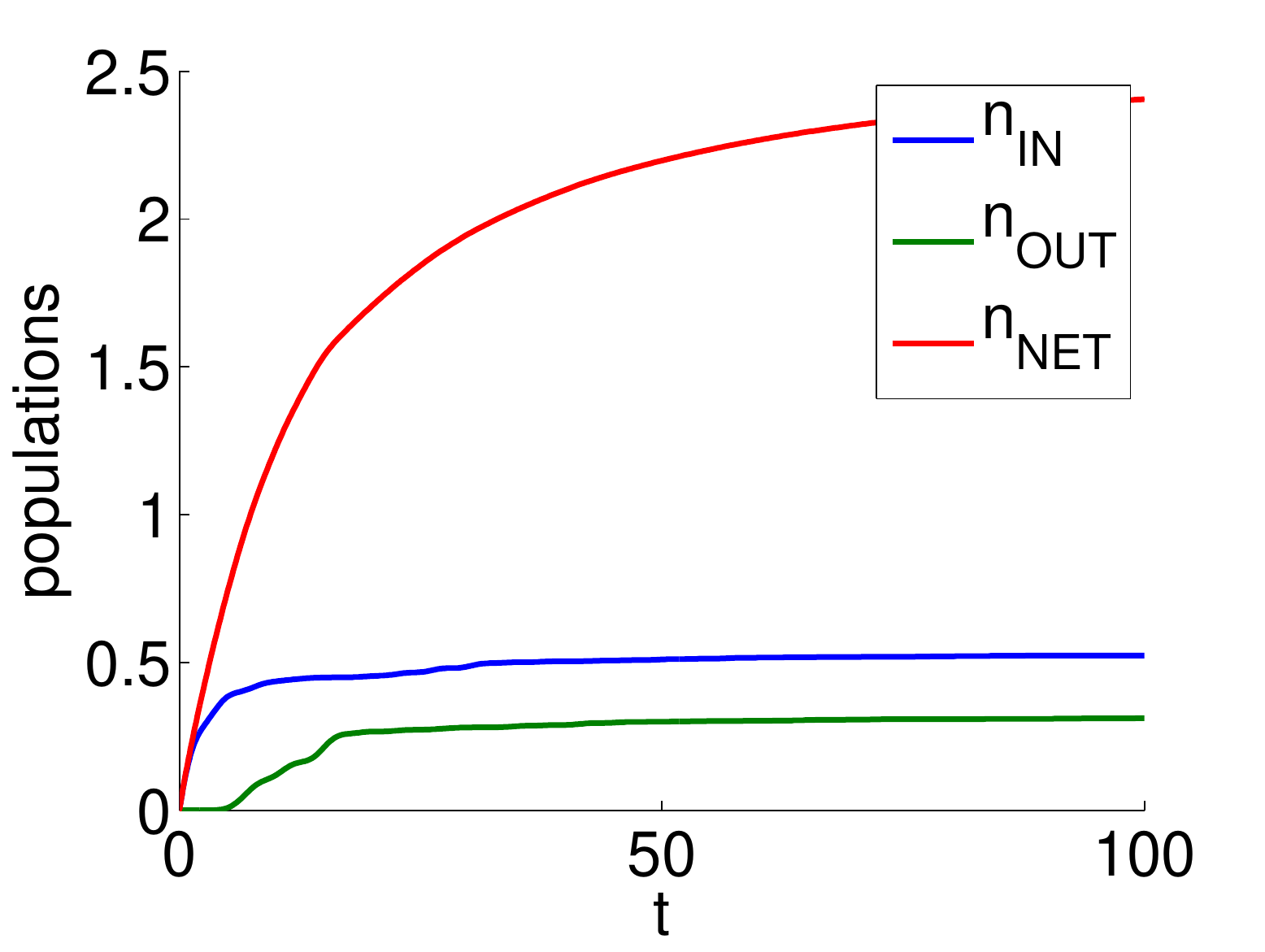}\includegraphics[width=4.5cm]{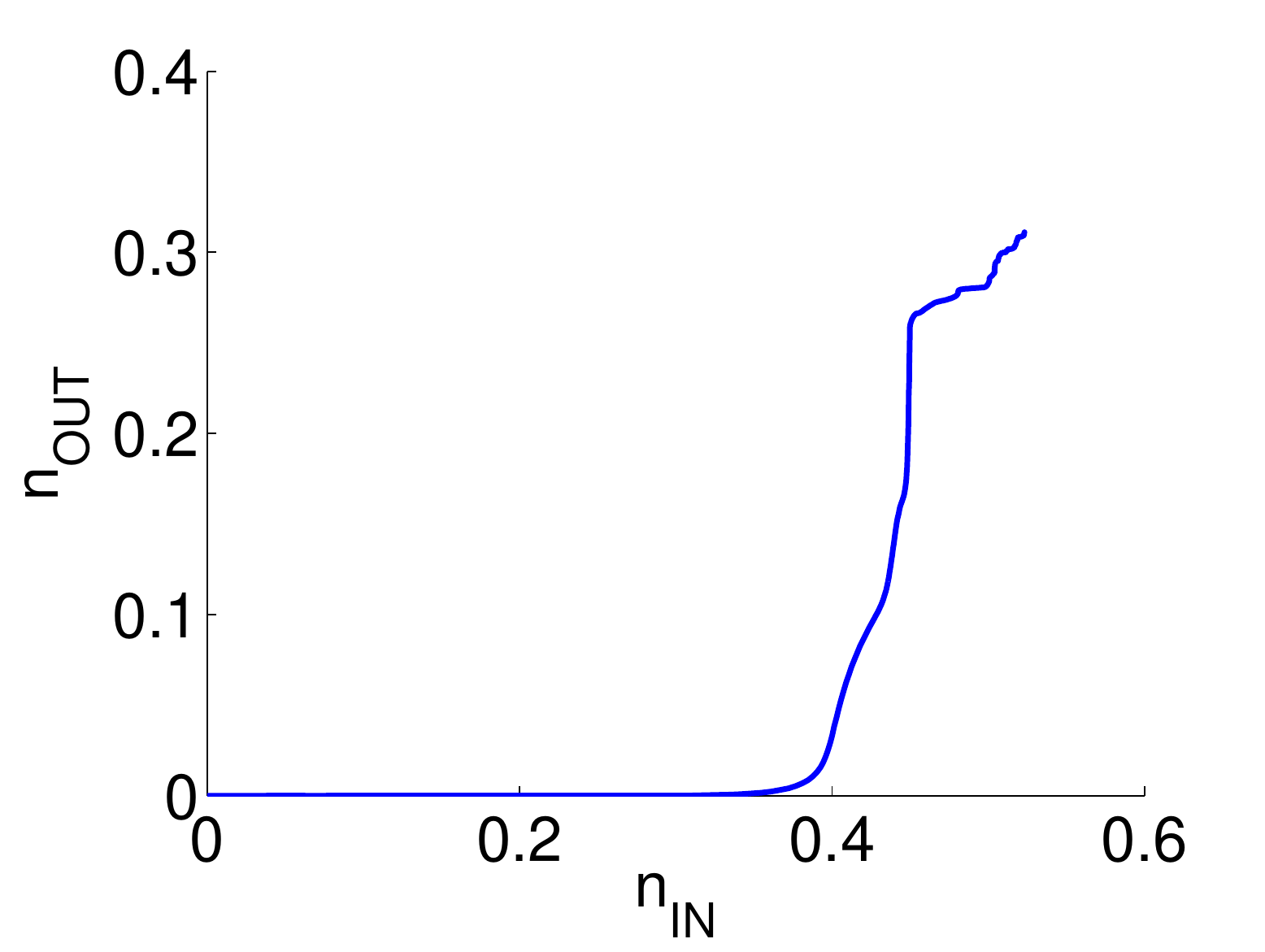} 
\par\end{centering}

\noindent \begin{centering}
\includegraphics[width=4.5cm]{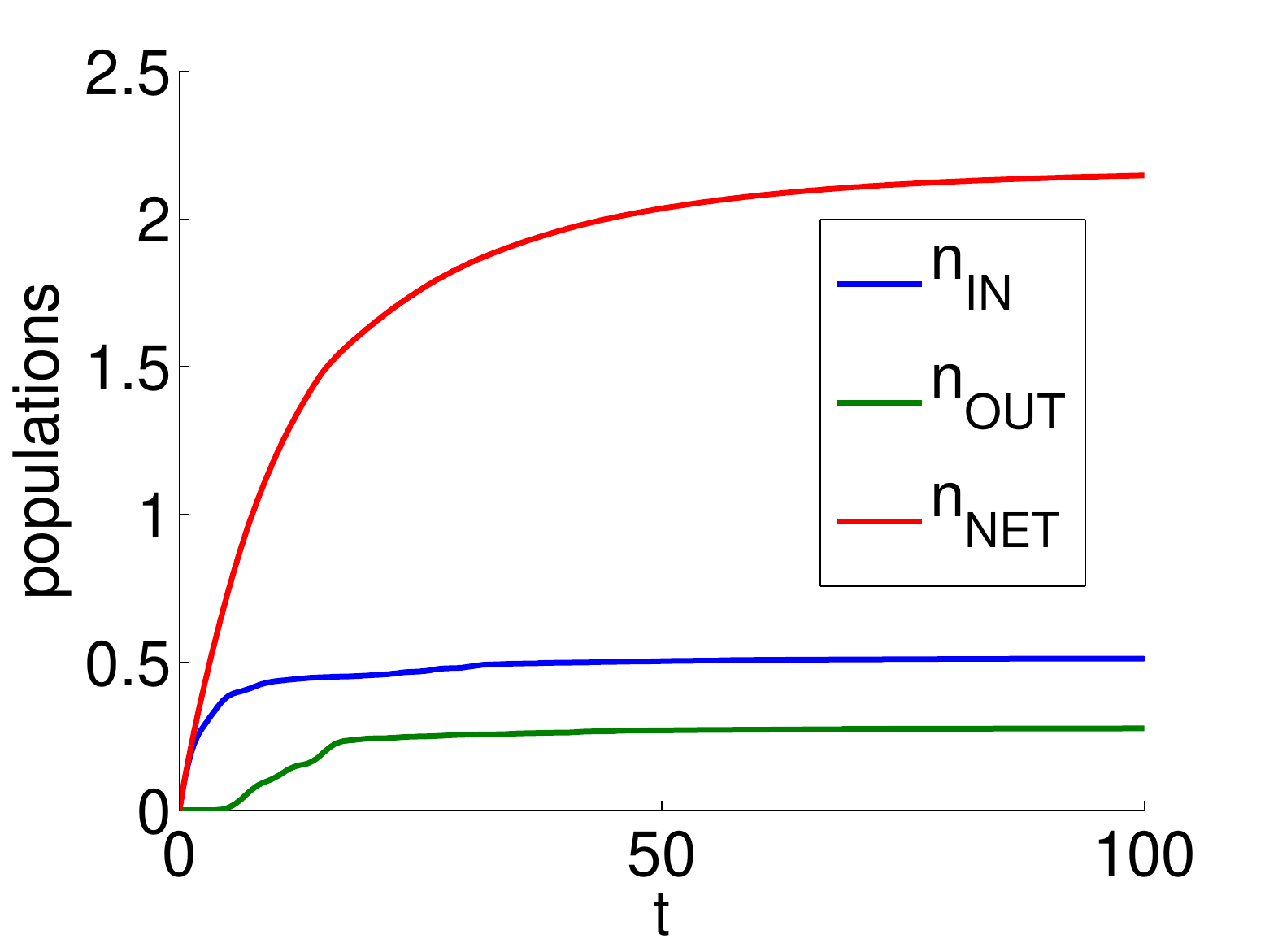}\includegraphics[width=4.5cm]{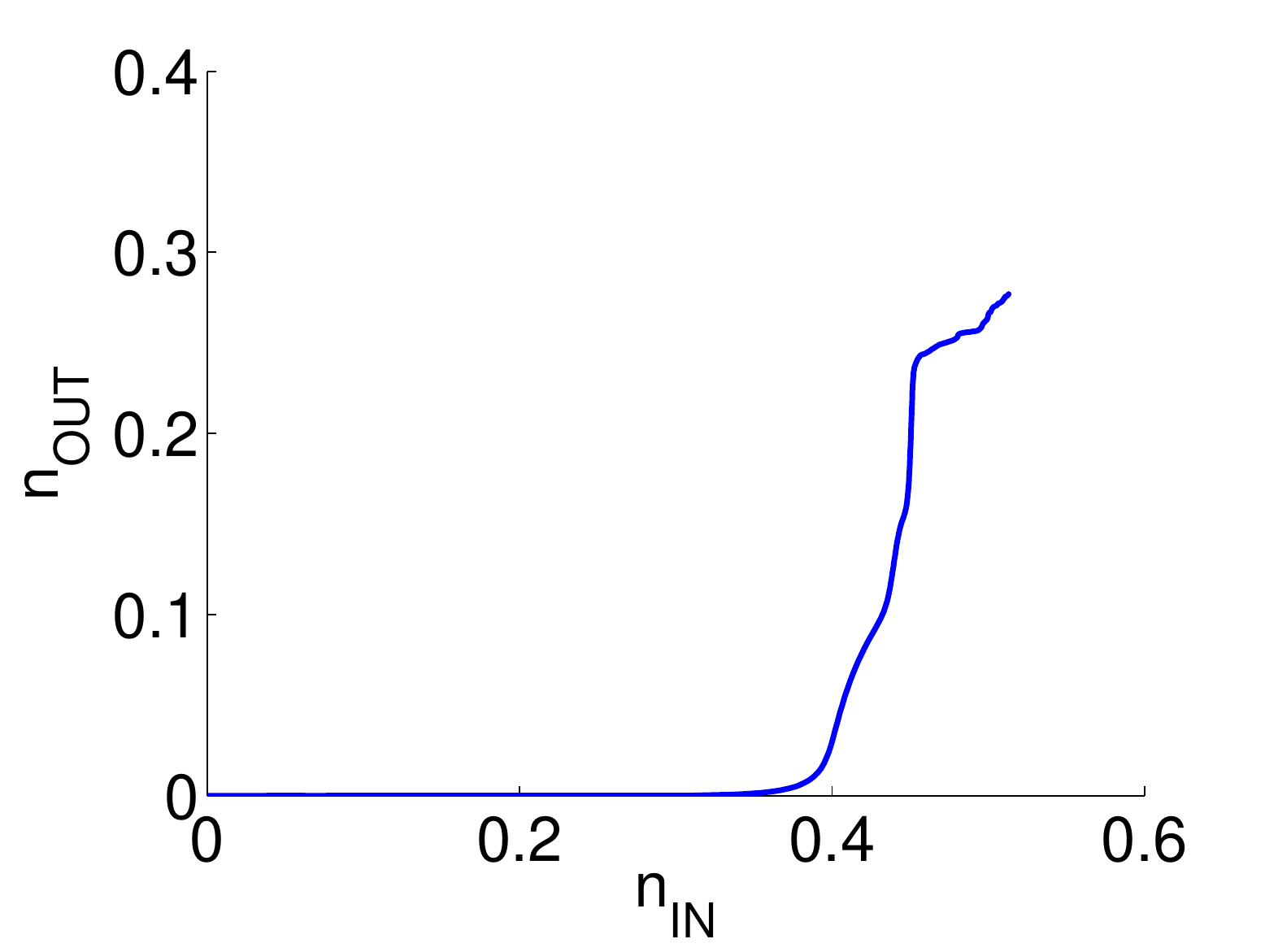} 
\par\end{centering}

\caption{$J=1$ and $\gamma_{\mathrm{in}}=0.2$, $\gamma_{\mathrm{out}}=0.3$.
First row panels: neither dissipation nor dephasing, and no static
noise. Second row panels: addition of diagonal static noise $\epsilon_{n}=J\cos\left(en\right)$.
Third row panels: static noise plus dissipation and dephasing $\gamma_{\mathrm{diss}}=\gamma_{\mathrm{deph}}=10^{-2}$.
\label{fig:ladder-results}}

\end{figure}

Here we want to show that the staircase effect, studied in Sec.~\ref{sec:staircase},
survives in more elaborate networks. We will study this effect in
the model depicted in Fig.~\ref{fig:ladder-picture}. The model consists
of an open chain of $N$ sites hopping coherently between nearest
neighbors, i.e.~the Hamiltonian is\[
H=\sum_{j=1}^{N-1}J\left(\sigma_{j}^{-}\sigma_{j+1}^{+}+\sigma_{j}^{+}\sigma_{j+1}^{-}\right).\]
 Particles are injected into the first site of the chain via a jump
operator $L_{\mathrm{in}}=\sqrt{\gamma_{\mathrm{in}}}\sigma_{1}^{+}$
and taken away at the last site via $L_{\mathrm{out}}=\sqrt{\gamma_{\mathrm{out}}}\sigma_{N}^{-}$.
On top of this basic framework we add different layers of complexity.
First we can add some static random diagonal noise, i.e.~we add site
dependent energies to the coherent part $H\rightarrow H+\sum_{j}\epsilon_{j}\mathfrak{n}_{j}$.
Second we can also include dissipation and dephasing acting on the
inner sites of the chain by adding the following superoperator: \emph{$\mathcal{L}_{\mathrm{noise}}=\sum_{j=2}^{N-1}\mathcal{L}_{L_{j,\mathrm{diss}}}+\mathcal{L}_{L_{j,\mathrm{deph}}}$}
($L_{j,\mathrm{diss}}=\sqrt{\gamma_{\mathrm{diss}}}\sigma_{j}^{-}$
and $L_{j,\mathrm{deph}}=\sqrt{\gamma_{\mathrm{deph}}}\mathfrak{n}_{j}$
as defined previously).

The picture that we have is the following. Through the coherent part
of the evolution, excitations travel in the chain in packets of quasiparticles
at velocities $v_{k}=O\left(J\right)$ ($k$ is a quasi-momentum label).
This introduces a lag timescale $T_{0}\approx L/v\sim O\left(L/J\right)$,
which is the time needed for an excitation to travel from one side
of the chain to the other. From Fig.~\ref{fig:ladder-results} (all
panels) we see that, when the population at the injection site increases,
the population at the expulsion site stays constant during this time-lag
$T_{0}$ and vice versa. Considering the central and lower panel of
Fig.~\ref{fig:ladder-results} we can appreciate how robust the effect
is with respect to various type of {}``perturbations''. The addition
of static random noise has the effect of localizing states and shuffling
the single-particle dispersion $\epsilon_{k}$. Both of these effects
destroy the picture of wavepackets traveling at constant velocity,
in that both the traveling times and the dispersion of the wave-packets
increase. Instead the addition of dissipation (and dephasing) to the
network, mostly has the effect of relaxing the system at a faster
rate. As long as the system has not relaxed the effect remains visible.
Comparatively, the presence of static coherent noise hinders the stair-case
effect more to dissipation and dephasing.

\section{Conclusions}

Inspired by the models which are recently being used to describe energy
transfer in photosynthetic pigments, we have identified and discussed
a few effects arising in quantum networks with coherent (Hamiltonian)
as well as incoherent (Lindblad) coupling between the nodes. For the
reader's sake we summarize here below these basic effects 
\begin{enumerate}
\item \emph{Congestion effect. }The incoherent transfer of excitations is
inversely proportional to the population in the reaction center. This
is due to the hard-core nature of the excitations that effectively
reduces the amplitude of the jump operator as the reaction center
fills. 
\item \emph{Asymptotic unitarity. } Coherent, unitary evolution may emerge
out of a dissipative, incoherent dynamics. This happens if states
which annihilate the incoherent part of the dynamics can be reached
during the time evolution. For this effect to be observable one needs
a separation of time-scales, $T_{\mathrm{relax}}\ll T_{\mathrm{diss}}$.
Such separation of time-scales does take place in some photosynthetic
systems e.g.~in the LH1-RC complexes present in purple bacteria. 
\item \emph{Staircase effect. }This effect refers to a situation in which
particles are injected incoherently, travel coherently along a given
chain, and then are expelled (or digested) at a certain rate at the
other end of the chain. The effect of the coherent part is to introduce
a time-scale $T_{0}=O\left(L/v\right)=O\left(L/J\right)$, ($L$ is
the system size, $v$ the velocity of excitations and $J$ is the
energy scale of the coherent network). $T_{0}$ is roughly the time
needed for the excitations to travel from one side of the chain to
the other. The peculiar feature emerging from the dynamic evolution,
is that when the population at the injection site increases, the population
at the expulsion site stays constant during this time-lag $T_{0}$
and vice versa. This effect results in a step-like behavior in the
parametric plot of the injection/extraction populations. 
\end{enumerate}
The effects we analyzed in this paper can be traced back to very simple
mechanisms displayed even by networks composed by only few qubits.
We provided analytical solutions for these toy models and showed numerical
evidence that these effects survive in more elaborated network such
as those modeling energy transfer in purple bacteria. Clearly, further
investigations are in order to establish the relevance of the elementary
calculations presented in this paper to the newborn field of quantum
biology.

The authors are sincerely in debt with Alán Aspuru-Guzik who played
a vital role in the early stage of this project. We also thank N.~Toby
Jacobson for a careful reading of the manuscript. P.Z.~acknowledges
support from NSF grants PHY-803304, DMR-0804914 and L.C.V.~acknowledges
support from European project COQUIT under FET-Open grant number 2333747.
\bibliographystyle{apsrev}
\bibliography{refs}

\end{document}